\documentclass[12pt]{article}
\usepackage{setspace}
\usepackage{longtable}
\usepackage{graphicx}

\usepackage{lscape}
\usepackage{rotating}

\usepackage[section]{placeins}
\usepackage[top=1in, left=1in, right=1in, bottom=1in]{geometry}
\usepackage[parfill]{parskip}	
\usepackage{graphicx}
\usepackage{caption}
\usepackage{subcaption}
\usepackage[export]{adjustbox}
\usepackage{amssymb}
\usepackage{epstopdf}
\usepackage{bbm}
\usepackage{bm}
\usepackage{amssymb}
\usepackage{amsmath}
\usepackage{amsfonts}

\usepackage{lscape}
\usepackage{enumitem}
\usepackage{rotating}
\usepackage{setspace}
\usepackage{threeparttable}
\usepackage{booktabs}
\usepackage[compact]{titlesec}
\usepackage{lmodern}
\fontfamily{lmtt}\selectfont
\usepackage[T1]{fontenc}

\newcommand*{\dbar}[1]{\overline{\overline{#1}}}

\usepackage[nocitation]{apacite}
\usepackage{natbib}
\bibpunct{(}{)}{;}{a}{}{,}
\usepackage{hyperref}
\usepackage{titling}
\usepackage{pifont}
\usepackage[capposition=top]{floatrow}
\newcommand{\subtitle}[1]{%
  \posttitle{%
    \par\end{center}
    \begin{center}\large#1\end{center}
    \vskip0.5em}%
}
\usepackage{amsthm}
\usepackage{amsmath} 
\usepackage{multirow}

\usepackage{multicol}
\usepackage{mathrsfs}

\newenvironment{psmallmatrix}
  {\left(\begin{smallmatrix}}  
  {\end{smallmatrix}\right)}

\newcommand{\norm}[1]{\left\lVert#1\right\rVert}

\newtheorem{theorem}{Theorem}
\newtheorem{lemma}{Lemma}
\newtheorem{proposition}{Proposition}

\def\ci{\perp\!\!\!\perp}

\renewcommand{\Pr}{\text{pr}}
\newcommand{\Var}{\text{var}}
\newcommand{\Cov}{\text{cov}}

\defcitealias{observational2021observational}{Observational Studies 2021}

\usepackage{mathrsfs}

\usepackage{color}

\usepackage{array}
\newcolumntype{C}[1]{>{\centering\arraybackslash}m{#1}}
\begin{document}
\pagestyle{plain}

\newtheoremstyle{mystyle}
{\topsep}
{\topsep}
{\it}
{}
{\bf}
{.}
{.5em}
{}
\theoremstyle{mystyle}
\newtheorem{assumptionex}{Assumption}
\newenvironment{assumption}
  {\pushQED{\qed}\renewcommand{\qedsymbol}{}\assumptionex}
  {\popQED\endassumptionex}
\newtheorem{assumptionexp}{Assumption}
\newenvironment{assumptionp}
  {\pushQED{\qed}\renewcommand{\qedsymbol}{}\assumptionexp}
  {\popQED\endassumptionexp}
\renewcommand{\theassumptionexp}{\arabic{assumptionexp}$'$}

\newtheorem{assumptionexpp}{Assumption}
\newenvironment{assumptionpp}
  {\pushQED{\qed}\renewcommand{\qedsymbol}{}\assumptionexpp}
  {\popQED\endassumptionexpp}
\renewcommand{\theassumptionexpp}{\arabic{assumptionexpp}$''$}

\newtheorem{assumptionexppp}{Assumption}
\newenvironment{assumptionppp}
  {\pushQED{\qed}\renewcommand{\qedsymbol}{}\assumptionexppp}
  {\popQED\endassumptionexppp}
\renewcommand{\theassumptionexppp}{\arabic{assumptionexppp}$'''$}

\renewcommand{\arraystretch}{1.3}

\newcommand{\argmin}{\mathop{\mathrm{argmin}}}
\makeatletter
\newcommand{\grande}{\bBigg@{2.25}}
\newcommand{\enorme}{\bBigg@{5}}

\newcommand{\blind}{0}

\newcommand{\tit}{\Large On the implied weights of linear regression for causal inference}

\if0\blind

{\title{\tit\thanks{For comments and conversations, we thank Peter Aronow, Eric Cohn, Avi Feller, David Hirshberg, Winston Lin, Bijan Niknam, Jamie Robins, Paul Rosenbaum, and Dylan Small. This work was supported through a Patient-Centered
Outcomes Research Institute (PCORI) Project Program Award (ME-2019C1-16172) and grants from the Alfred P. Sloan Foundation (G-2018-10118, G-2020-13946).}}
\author{Ambarish Chattopadhyay\thanks{Department of Statistics, Harvard University, 1 Oxford Street
Cambridge, MA 02138; email: \url{ambarish_chattopadhyay@g.harvard.edu}.} \and Jos\'{e} R. Zubizarreta\thanks{Departments of Health Care Policy, Biostatistics, and Statistics, Harvard University, 180 A Longwood Avenue, Office 307-D, Boston, MA 02115; email: \url{zubizarreta@hcp.med.harvard.edu}.}
}

\date{} 

\maketitle
}\fi

\if1\blind
\title{\bf \tit}
\date{} 
\maketitle
\fi

\begin{abstract}
A basic principle in the design of observational studies is to approximate the randomized experiment that would have been conducted under ideal circumstances. 
In practice, linear regression models are commonly used to analyze observational data and estimate causal effects.
How do linear regression adjustments in observational studies emulate key features of randomized experiments, such as covariate balance, self-weighted sampling, and study representativeness?
In this paper, we provide answers to this and related questions by analyzing the implied (individual-level data) weights of various linear regression methods, bringing new insights at the intersection of regression modeling and causal inference.
We derive new closed-form expressions of these implied weights and examine their properties in finite and large samples.
Among others, in finite samples we characterize the implied target population of linear regression and in large samples demonstrate the multiply robust properties of regression estimators from the perspective of their implied weights.
We show that the implied weights of general regression methods can be equivalently obtained by solving a convex optimization problem.
This equivalence allows us to bridge ideas from the regression modeling and causal inference literatures.
As a result, we propose novel regression diagnostics for causal inference that are part of the design stage of an observational study.
We implement the weights and diagnostics in the new \texttt{lmw} package for \texttt{R}.
\end{abstract}


\begin{center}
\end{center}
\clearpage
\doublespacing

\singlespacing
\pagebreak
\tableofcontents
\pagebreak
\doublespacing


\section{Introduction}
\label{sec_introduction}

\subsection{Regression and experimentation}

In a landmark paper in 1965, Cochran recommended that ``the planner of an observational study should always ask himself the question, `How would the study be conducted if it were possible to do it by controlled experimentation?''' (\citealt{cochran1965planning}).
Some key features of randomized experiments are: (a) covariate balance, i.e., pre-treatment variables are balanced in expectation; (b) study representativeness, i.e., the target population for inference is the experimental sample itself or a broader population under a known sample selection mechanism; (c) self-weighted sampling, i.e., randomization produces unweighted samples of treated and control units, and analyses that impose differential weights on the units reduce the efficiency of estimators; and (d) sample boundedness, i.e., since covariates are balanced in expectation, covariate adjustments and effect estimates are mostly an interpolation and not an extrapolation beyond the support of the observed data.
At present, linear regression models are extensively used to analyze observational data and estimate average causal effects.
But to what extent does regression emulate these key features of a randomized experiment?
More concretely, how does regression adjust for or balance the covariates included as regressors in the model? 
What is the population that regression adjustments actually target?
And what is the connection between regression and other methods for statistical adjustment, such as matching and weighting?

In this paper, we answer these and related questions.
In particular, we show how linear regression acts on the individual-level data to approximate a randomized experiment and produce an average treatment effect estimate.
To this end, we examine how regression implicitly weights the treatment and control individual observations by finding its implied weights.

\subsection{Contribution and related works}

In this paper, we derive and analyze the implied weights of various linear regression estimators.
We obtain new closed-form, finite sample expressions of the weights. 
The regression estimators we consider include: (1) uni-regression imputation (URI), which arguably is the most common regression estimator in practice and is equivalent to estimating the coefficient of the treatment indicator in a linear regression model on the covariates and the treatment without including their interactions, (2) multi-regression imputation (MRI) or g-computation, which is equivalent to estimating the coefficient of the treatment indicator in a similar model including interactions, (3) weighted least squares analogs of URI and MRI, and (4) augmented inverse probability weighting (\citealt{robins1994estimation}).

In relation to URI, previous works have obtained weighting representations of the regression estimand (\citealt{angrist2008mostly}, Chapter 3) or the regression estimators using asymptotic approximations (\citealt{aronow2016does}).
Our work differs from them in that we provide a weighting representation of the URI estimator in closed forms and in finite samples.
This representation clarifies how URI acts on each individual-level observation in the study sample to produce an average treatment effect estimate. 
In relation to MRI, previous works have derived closed-form expressions of the MRI weights in specific settings such as univariate regression (\citealt{imbens2015matching}), synthetic controls (\citealt{abadie2015comparative}, \citealt{ben2021augmented}), and regression discontinuity designs   (\citealt{gelman2018high}). Some of these results are analogous to those of regression estimation in sample survey (\citealt{huang1978nonnegative}, \citealt{deville1992calibration}). 
Our work builds on and generalizes these important contributions to general multivariate MRI estimators for several estimands of interest. 

We derive the properties of the implied weights and the corresponding estimators in both finite and large sample regimes. 
In finite samples, we analyze the weights in terms of (a) covariate balance, (b) study representativeness, (c) dispersion, (d) sample-boundedness, and (e) optimality. 
In particular, we characterize the implied target population of linear regression in finite samples, which to our knowledge has not been done in causal inference. 
Moreover, using the implied weights, we provide an alternative view of regression adjustments from the standpoint of mathematical optimization, and in turn, connect regression to modern matching and weighting methods in causal inference. 
In large samples, we show that under certain conditions URI and MRI are equivalent to inverse probability weighting. 
Leveraging this equivalence, we show that  the URI and MRI estimators are multiply robust. 
Our results generalize previous results on double robustness for regression estimators in survey sampling (\citealt{robins2007comment}) and causal inference (\citealt{kline2011oaxaca}) to multiple robustness and other types of regression estimators such as URI. 


Finally, this implied weighting framework allows us to propose new regression diagnostics for causal inference.
These diagnostics assess covariate balance, model extrapolation, dispersion of the weights and effective sample size, and influence of a given observation on an estimate of the average treatment effect. 
Conventionally, regression adjustments are viewed as part of the analysis stage of an observational study, but as we discuss in this paper, they can be conducted as part of the design stage \citep{rubin2008objective}. In this sense,
our first three proposed diagnostics can be regarded as design-based diagnostics for regression adjustments in observational studies.

\section{Notation, estimands, and assumptions}
\label{sec_notation} 

We operate under the potential outcomes framework for causal inference (\citeauthor{neyman1923application} 1923, 1990, \citealt{rubin1974estimating}) and consider a sample of $n$ units randomly drawn from a population. 
For each unit $i = 1, ..., n$, $Z_i$ is a treatment assignment indicator with $Z_i = 1$ if the unit is assigned to treatment and $Z_i = 0$ otherwise; $\bm{X}_i \in \mathbb{R}^k$ is a vector of observed covariates; and $Y^{\text{obs}}_i$ is the observed outcome variable. Under the Stable Unit Treatment Value Assumption (SUTVA; \citealt{rubin1980randomization}), let $\{Y_i(1), Y_i(0) \}$ be the potential outcomes under treatment and control, respectively, where only one of them is observed in the sample: $Y^{\text{obs}}_i = Z_i Y_i(1) + (1-Z_i)Y_i(0)$.

We focus on estimating the Average Treatment Effect (ATE), defined as $\text{ATE} = {E}\{Y_i(1) - Y_i(0)\}$, and the Average Treatment Effect on the Treated (ATT), given by $\text{ATT} = {E}\{Y_i(1) - Y_i(0)\mid Z_i=1\}$. 
We also consider the Conditional Average Treatment Effect (CATE). 
The CATE for a population with a fixed covariate profile $\bm{x}^* \in \mathbb{R}^k$ is given by $\text{CATE}(\bm{x}^*) = {E}\{Y_i(1) - Y_i(0)\mid \bm{X}_i = \bm{x}^*\}$; i.e., the CATE is the ATE in the subpopulation of units with covariate vector equal to $\bm{x}^*$. 

For identification of these estimands, we assume that the treatment assignment satisfies the unconfoundedness and positivity assumptions: $Z_i \ci \{Y_i(0),Y_i(1)\} \mid \bm{X}_i$ and $0<\Pr(Z_i = 1\mid \bm{X}_i = \bm{x})<1$ for all $\bm{x} \in \text{supp}(\bm{X}_i)$, respectively \citep{rosenbaum1983central}.
For conciseness, we adopt the following additional notation.
Denote the conditional mean functions of the potential outcomes under treatment and control as $m_1(\bm{x}) = {E}\{Y_i(1)\mid \bm{X}_i=\bm{x}\}$ and $m_0(\bm{x}) = {E}\{Y_i(0)\mid \bm{X}_i=\bm{x}\}$, respectively.
Let $n_t = \sum_{i=1}^{n}Z_i$ and $n_c = \sum_{i=1}^{n}(1-Z_i)$ be the treatment and control group sizes. 
Write $\bm{\underline{X}}$ for the $n\times k$ matrix of covariates in the full sample that pools the treatment and control groups, and let $\bar{\bm{X}}_t = \sum_{i:Z_i=1} \bm{X}_i/n_t$ and $\bar{\bm{X}}_c = \sum_{i:Z_i=0} \bm{X}_i/n_c$. 
The average of the covariate vectors $\bm{X}_i$ in the full sample is given by $\bar{\bm{X}} = n^{-1}(n_t \bar{\bm{X}}_t + n_c \bar{\bm{X}}_c)$. 
Also, let $\bm{S}_t = \sum_{i:Z_i=1}(\bm{X}_i - \bar{\bm{X}}_t)(\bm{X}_i - \bar{\bm{X}}_t)^\top$ and $\bm{S}_c = \sum_{i:Z_i=0}(\bm{X}_i - \bar{\bm{X}}_c)(\bm{X}_i - \bar{\bm{X}}_c)^\top$ be the scaled covariance matrices in the treatment and control group respectively. Throughout the paper, we assume that $\bm{S}_t$ and $\bm{S}_c$ are invertible. 
Finally, let $\bar{Y}_t$ and $\bar{Y}_c$ be the mean observed outcomes in the treatment and control group, respectively.

\section{Implied weights of linear regression}
\label{sec_implied}

A widespread approach to estimate the ATE goes as follows.
On the entire sample, use ordinary least squares (OLS) to fit a linear regression model of the observed outcome $Y^{\text{obs}}_i$ on the baseline covariates $\bm{X}_i$ and the treatment indicator $Z_i$, and compute the coefficient associated with $Z_i$ (see, e.g., Chapter 3 of \citealt{angrist2008mostly} and Section 12.2.4 of \citealt{imbens2015causal}). 
Under mean unconfoundedness, this approach can be motivated by the structural model $Y_i(z) = \beta_0 + \bm{\beta}^\top_1 \bm{X}_i + \tau z + \epsilon_{iz} , \hspace{0.1cm}{E}(\epsilon_{iz}\mid \bm{X}_i) = 0$, $z \in \{0,1\}$. 
Here the CATE is constant and equal to $\tau$ across the space of the covariates; i.e., $m_1(\bm{x}) - m_0(\bm{x}) = \tau$. 
Thus $\text{ATE} = {E}\{m_1(\bm{X}_i) - m_0(\bm{X}_i)\} = \tau$, and by unconfoundedness, ${E}\{Y_i(z)\mid \bm{X}_i = \bm{x}\} = {E}\{Y^{\text{obs}}_i\mid \bm{X}_i = \bm{x},Z_i = z\} = \beta_0 + \bm{\beta}^\top_1 \bm{x} + \tau z$.
By standard linear model theory, if the model for $Y_i(z)$ is correct, then the OLS estimator $\hat{\tau}^{{\scriptscriptstyle \text{OLS}}}$ is the best linear unbiased and consistent estimator for the ATE. 

Now, since $\text{ATE} = {E}\{ m_1(\bm{X}_i) - m_0(\bm{X}_i)\}$, a natural way to estimate this quantity is to compute its empirical analog $n^{-1}\sum_{i=1}^{n} \{ {m}_1(\bm{X}_i) - {m}_0(\bm{X}_i) \}$. Therefore, a broad class of imputation estimators of the ATE has the form
$\widehat{\text{ATE}} = n^{-1}\sum_{i=1}^{n} \{ \hat{m}_1(\bm{X}_i) - \hat{m}_0(\bm{X}_i) \}$, where $\hat{m}_1(\bm{x})$ and $\hat{m}_0(\bm{x})$ are some estimators of ${m}_1(\bm{x})$ and ${m}_0(\bm{x})$, respectively. 
Such imputation estimators are popular in causal inference (see, e.g., Chapter 13 of \citealt{hernan2020causal}).
 
Clearly, $\hat{\tau}^{{\scriptscriptstyle \text{OLS}}}$ is also an imputation estimator. 
Henceforth, we term this approach uni-regression imputation (URI), because the potential outcomes are imputed using a single (uni) regression model.
In Proposition \ref{prop_uri} we show that $\hat{\tau}^{{\scriptscriptstyle \text{OLS}}}$ can be represented as a difference of weighted means of the treated and control outcomes.
We also provide closed form expressions for the implied regression weights. 

\begin{proposition}
\label{prop_uri}
The URI estimator of the ATE can be expressed as $\hat{\tau}^{{\scriptscriptstyle \text{OLS}}} = \sum_{i:Z_i=1}w^{{\scriptscriptstyle \text{URI}}}_i Y^{\text{obs}}_i - \sum_{i:Z_i=0}w^{{\scriptscriptstyle \text{URI}}}_i Y^{\text{obs}}_i$ where $w^{{\scriptscriptstyle \text{URI}}}_i = n^{-1}_t + n n^{-1}_c (\bm{X}_i  -\bar{\bm{X}}_t)^\top (\bm{S}_t + \bm{S}_c)^{-1} (\bar{\bm{X}} - \bar{\bm{X}}_t)$ for each unit in the treatment group and $w^{{\scriptscriptstyle \text{URI}}}_i = n^{-1}_c + n n^{-1}_t (\bm{X}_i  -\bar{\bm{X}}_c)^\top (\bm{S}_t + \bm{S}_c)^{-1} (\bar{\bm{X}} - \bar{\bm{X}}_c)$ for each unit in the control group.
Moreover, within each group the weights add up to one, $\sum_{i:Z_i=0} w^{{\scriptscriptstyle \text{URI}}}_i =1$ and $\sum_{i:Z_i=1} w^{{\scriptscriptstyle \text{URI}}}_i =1$.
\end{proposition} 
A proof for Proposition \ref{prop_uri}, as well as for all the other results in the paper, is presented in the Supplementary Material.
According to Proposition \ref{prop_uri}
the regression estimator $\hat{\tau}^{{\scriptscriptstyle \text{OLS}}}$ is a H\'{a}jek estimator with weights $w^{{\scriptscriptstyle \text{URI}}}_i$. 
Furtheremore, we see that the URI weights depend on the treatment indicators and the covariates but not on the observed outcomes. 
Therefore, although typical software implementations of URI require the outcomes and simultaneously adjust for the covariates and produce effect estimates, the weighting representation in Proposition \ref{prop_uri} shows that the linear regression model can be ``fit'' without the outcomes.
In other words, using \cite{rubin2008objective}'s classification of the stages of an observational study, the URI weights can be obtained as a part of the `design stage' of the study, as opposed to its `analysis stage,' helping to preserve the objectivity of the study and bridge ideas from matching and weighting to regression modeling.

Another type of imputation estimator obtains $\hat{m}_1(\bm{x})$ and $\hat{m}_0(\bm{x})$ by fitting two separate linear regression models on the treatment and control samples, given by $Y^{\text{obs}}_i = \beta_{0t} + \bm{\beta}^\top_{1t}\bm{X}_i + \epsilon_{it}$ and $Y^{\text{obs}}_i = \beta_{0c} + \bm{\beta}^\top_{1c}\bm{X}_i + \epsilon_{ic}$, respectively. 
This approach is more flexible than the former since it allows for treatment effect modification. 
In particular, under mean unconfoundedness, the conditional average treatment effect is linear in the covariates, i.e.,  $\text{CATE}(\bm{x}) = m_1(\bm{x}) - m_0(\bm{x}) = (\beta_{0t} - \beta_{0c}) + (\bm{\beta}_{1t} - \bm{\beta}_{1c})^\top \bm{x}$. 
We call this approach multi-regression imputation (MRI). 
Clearly, MRI and URI are equivalent if the model used in the URI approach includes all possible interaction terms between the treatment indicator and the mean-centered covariates.
With the MRI approach, it is convenient to estimate a wide range of estimands, including the ATE, ATT, and the CATE.
The following proposition shows the implied form of weighting of the treated and control units under the MRI approach.

\begin{proposition}
\label{prop_mri}
The MRI estimators of the ATE, ATT, and CATE can be expressed as
\begin{enumerate}[label=(\alph*)]
\item $\widehat{\text{ATE}} = \sum_{i:Z_i = 1}w^{{\scriptscriptstyle \text{MRI}}}_i(\bar{\bm{X}}) Y^{\text{obs}}_i - \sum_{i:Z_i = 0}w^{{\scriptscriptstyle \text{MRI}}}_i(\bar{\bm{X}}) Y^{\text{obs}}_i$,

\item $\widehat{\text{ATT}} = \bar{Y}_t - \sum_{i:Z_i = 0}w^{{\scriptscriptstyle \text{MRI}}}_i(\bar{\bm{X}}_t) Y^{\text{obs}}_i$, and

\item $\widehat{\text{CATE}}(\bm{x}^*) = \sum_{i:Z_i = 1}w^{{\scriptscriptstyle \text{MRI}}}_i(\bm{x}^*)Y^{\text{obs}}_i - \sum_{i:Z_i = 0}w^{{\scriptscriptstyle \text{MRI}}}_i(\bm{x}^*) Y^{\text{obs}}_i$,
\end{enumerate}
where $w^{{\scriptscriptstyle \text{MRI}}}_i(\bm{x}) = n^{-1}_t + (\bm{X}_i -\bar{\bm{X}}_t)^\top \bm{S}_t^{-1} (\bm{x} - \bar{\bm{X}}_t)$ if unit $i$ is in the treatment group and $w^{{\scriptscriptstyle \text{MRI}}}_i(\bm{x}) = n^{-1}_c + (\bm{X}_i -\bar{\bm{X}}_c)^\top \bm{S}_c^{-1} (\bm{x} - \bar{\bm{X}}_c)$ if unit $i$ is in the control group.
Moreover, $\sum_{i:Z_i=1}w^{{\scriptscriptstyle \text{MRI}}}_i(\bm{x}) = \sum_{i:Z_i=0}w^{{\scriptscriptstyle \text{MRI}}}_i(\bm{x}) = 1$ for all $\bm{x} \in \mathbb{R}^k$.
\end{proposition}

We observe that, similar to the URI weights, the MRI weights do not depend on the outcomes and hence can be part of the design stage of the study. 
Propositions \ref{prop_uri} and \ref{prop_mri} highlight how the implied weights depart from uniform weights as a function of covariate balance before adjustments. In particular, both URI and MRI weights become uniform if the covariates in the treatment and control groups are exactly mean balanced a priori. 
These weighting expressions can show when a particular observation has a large impact on the analysis via its implied weight (see Section \ref{sec_regression} for related diagnostics). 

Finally, propositions \ref{prop_uri} and \ref{prop_mri} imply that $w^{{\scriptscriptstyle \text{URI}}}_i = w^{{\scriptscriptstyle \text{MRI}}}_i(\bm{x^*})$, where $\bm{x^*} = \bm{S}_c(\bm{S}_t + \bm{S}_c)^{-1} \bar{\bm{X}}_t + \bm{S}_t (\bm{S}_t + \bm{S}_c)^{-1} \bar{\bm{X}}_c  $. This means that the URI weights are a special case of the MRI weights, where we impute the potential outcomes of a unit with $\bm{x} = \bm{x^*}$. 
In particular, a sufficient condition for $w^{{\scriptscriptstyle \text{URI}}}_i = w^{{\scriptscriptstyle \text{MRI}}}_i(\bar{\bm{X}})$ is that $n_t\bm{S}_t = n_c\bm{S}_c$, which holds if the treatment groups are of equal size and have the same sample covariance matrix. 
Indeed, another sufficient condition for the weights to be equal is that $\bar{\bm{X}}_t = \bar{\bm{X}}_c$, which implies that both weights are uniform.

\section{Properties of the implied weights}
\label{sec_properties}

\subsection{Finite sample properties}
\label{sec_finite}

In this section, we study the finite sample properties of the implied weights in regards to: (a) covariate balance, (b)  representativeness of the weighted sample, (c) dispersion or variability, (d) extrapolation, and (e) optimality from a mathematical programming standpoint.
The following proposition summarizes these properties for both the URI and the MRI weights for the ATE estimation problem. Henceforth, we denote the MRI weights for the ATE as $w^{\scriptscriptstyle \text{MRI}}_i$.
\begin{proposition}
\label{prop_finite_sample_properties}
\begin{enumerate}[label =(\alph*)]
\item[]
\item {Balance}: The URI and MRI weights exactly balance the means of the covariates included in the model, with respect to different profiles $\bm{X}^{*{\scriptscriptstyle \text{URI}}}$ and $\bm{X}^{*{\scriptscriptstyle \text{MRI}}}$:
$
\sum_{i:Z_i=1} w^{{\scriptscriptstyle \text{URI}}}_i \bm{X}_i = \sum_{i:Z_i=0} w^{{\scriptscriptstyle \text{URI}}}_i \bm{X}_i = \bm{X}^{*{\scriptscriptstyle \text{URI}}}, \quad \sum_{i:Z_i=1} w^{{\scriptscriptstyle \text{MRI}}}_i \bm{X}_i = \sum_{i:Z_i=0} w^{{\scriptscriptstyle \text{MRI}}}_i \bm{X}_i = \bm{X}^{*{\scriptscriptstyle \text{MRI}}}.
$

\item {Representativeness}: With the URI and MRI weights, the covariate profiles are
$
\bm{X}^{*{\scriptscriptstyle \text{URI}}} = \bm{S}_c(\bm{S}_t + \bm{S}_c)^{-1} \bar{\bm{X}}_t + \bm{S}_t (\bm{S}_t + \bm{S}_c)^{-1} \bar{\bm{X}}_c$ and $\bm{X}^{*{\scriptscriptstyle \text{MRI}}} = \bar{\bm{X}},
$ respectively.

\item {Dispersion}: The variances of the URI weights in the treatment and control groups are given by
$n^2 n^{-1}_t n^{-2}_c (\bar{\bm{X}} - \bar{\bm{X}}_t)^\top (\bm{S}_t + \bm{S}_c)^{-1}\bm{S}_t (\bm{S}_t + \bm{S}_c)^{-1} (\bar{\bm{X}} - \bar{\bm{X}}_t)
$ and $n^2n^{-1}_c n^{-2}_t (\bar{\bm{X}} - \bar{\bm{X}}_c)^\top (\bm{S}_t + \bm{S}_c)^{-1}\bm{S}_c (\bm{S}_t + \bm{S}_c)^{-1} (\bar{\bm{X}} - \bar{\bm{X}}_c)
$, respectively. 
Similarly, the variances of the MRI weights in the treatment and control groups are
$n^{-1}_t (\bar{\bm{X}} - \bar{\bm{X}}_t)^\top \bm{S}_t^{-1}(\bar{\bm{X}} - \bar{\bm{X}}_t)
$ and $n^{-1}_c (\bar{\bm{X}} - \bar{\bm{X}}_c)^\top \bm{S}_c^{-1}(\bar{\bm{X}} - \bar{\bm{X}}_c)$, respectively.

\item {Extrapolation}: The URI and MRI weights can both take negative values and produce average treatment effect estimators that are not sample bounded.

\item {Optimality}: The URI and MRI weights are the weights of minimum variance that add up to one and satisfy the corresponding covariate balance constraints in (a).

\end{enumerate}
\end{proposition}

One of our motivating questions was, to what extent does regression emulate the key features of a randomized experiment?
Proposition \ref{prop_finite_sample_properties} provides answers to this question.
Part (a) says that linear regression, both in its URI and MRI variants, exactly balances the means of the covariates included in the model, but with respect to different profiles, $\bm{X}^{*{\scriptscriptstyle \text{URI}}}$ and $\bm{X}^{*{\scriptscriptstyle \text{MRI}}}$.
Part (b) provides closed form expressions for these profiles, which in turn characterize the implied target populations of regression.
While MRI exactly balances the means of the covariates at the overall study sample mean, URI balances them elsewhere. In this sense, the URI weights may distort the structure of the original study sample.
Under linearity of the potential outcomes models, if treatment effects are homogeneous, both URI and MRI produce unbiased estimators of the ATE. However, if treatments effects are heterogeneous, then even under linearity, the URI estimator is biased for the ATE, whereas the MRI estimator is unbiased. See the Supplementary Material for details.

Part (c) characterizes the variances of the weights. 
For instance, the variance of both the URI and MRI weights in the treatment group are a scaled distance between $\bar{\bm{X}}$ and  $\bar{\bm{X}}_t$, multiplied by positive definite matrices. 
Since $(\bar{\bm{X}} - \bar{\bm{X}}_t)  = n^{-1}n_c (\bar{\bm{X}}_c - \bar{\bm{X}}_t)$, the variance of both the URI and MRI weights can also be interpreted as a distance similar to the Mahalanobis distance between the treatment and the control groups.
Therefore, for fixed $\bm{S}_t$ and $\bm{S}_c$, using URI or MRI on an a priori well-balanced sample 
will lead to weights that are less variable than that on an imbalanced sample. The variance of the weights is important because it directly impacts the variance of a weighted estimator.

Part (d) establishes that both the URI and MRI weights can take negative values, so the corresponding estimators are not sample bounded in the sense of \cite{robins2007comment} and their estimates can lie outside the support, or convex hull, of the observed outcome data. 
This property has been noted in instances of MRI, for example, in simple regression estimation of the ATT by \cite{imbens2015matching} and in synthetic control settings by \cite{abadie2015comparative}, but not in general. 
We refer the reader to Section \ref{sec_extrapolation} for a discussion on the implications of this property.

Finally, part (e) groups these results and states that the URI and MRI weights are the least variable weights that add up to one and exactly balance the means of the covariates included in the models with respect to given covariate profiles. 
This result helps to establish a connection between the implied linear regression weights and existing matching and weighting methods. For example, part (e) shows that URI and MRI can be viewed as weighting approaches with moment-balancing conditions on the weights, such as entropy balancing (\citealt{hainmueller2012balancing}) and the stable balancing weights (\citealt{zubizarreta2015stable}). 
We also note the connection of the implied weights to matching approaches, e.g., cardinality matching \citep{zubizarreta2014matching} where the weights are constrained to be constant integers representing a matching ratio and an explicit assignment between matched units.
In connection to sample surveys, Part (e) establishes URI and MRI as two-step calibration weighting methods (\citealt{deville1992calibration}), where the weights are calibrated separately in the treatment and control groups. 
See the Supplementary Material for results analogous to Proposition \ref{prop_finite_sample_properties} when the estimand is the ATT or the CATE($\bm{x}$).

\subsection{Asymptotic properties}
\label{sec_asymptotic_properties}


In this section, we study the large-sample behavior of the URI and MRI weights and their associated estimators.
This analysis reveals a connection between regression imputation and inverse probability weighting (IPW).
In particular, we show that under a given functional form for the true propensity score model, the MRI weights converge pointwise to the corresponding true inverse probability weights.
Moreover, the convergence is uniform if the $L_2$ norm of the covariate vector is bounded over its support. 
Theorem \ref{thm_mri_convergence} formalizes this result for the ATE estimation problem.
An analogous result holds for the ATT. 

\begin{theorem}
\label{thm_mri_convergence}
Suppose we wish to estimate the ATE.
Let $w^{{\scriptscriptstyle \text{MRI}}}_{\bm{x}}$ be the MRI weight of a unit with covariate vector $\bm{x}$.  Then
\begin{enumerate}[label = (\alph*)]

\item For each treated unit, $n w^{{\scriptscriptstyle \text{MRI}}}_{\bm{x}} \xrightarrow[n \to \infty]{P} 1/e(\bm{x})$ for all $\bm{x} \in \text{supp}(\bm{X}_i)$ if and only if the propensity score is an inverse linear function of the covariates; i.e., $e(\bm{x}) = 1/(\alpha_0 + \bm{\alpha}^\top_1 \bm{x})$, $\alpha_0 \in \mathbb{R}$, $\bm{\alpha}_1 \in \mathbb{R}^k$. Moreover, if $\sup_{\bm{x} \in \text{supp}(\bm{X}_i)}\norm{\bm{x}}_2 < \infty$, then $\sup_{\bm{x} \in \text{supp}(\bm{X}_i)}\mid nw^{{\scriptscriptstyle \text{MRI}}}_{\bm{x}} - \{1/e(\bm{x})\}\mid  \xrightarrow[n \to \infty]{P} 0$.

\item Similarly, for each control unit, $n w^{{\scriptscriptstyle \text{MRI}}}_{\bm{x}} \xrightarrow[n \to \infty]{P} 1/\{1-e(\bm{x})\}$ if and only if $1-e(\bm{x})$ is an inverse linear function of the covariates, and the convergence is uniform if $\sup_{\bm{x} \in \text{supp}(\bm{X}_i)}\norm{\bm{x}}_2 < \infty$.

\end{enumerate}
\end{theorem} 

Theorem \ref{thm_mri_convergence} says that, by fitting a linear regression model of the outcome in the treatment group, we implicitly estimate the propensity score.
Moreover, it says that if the true propensity score model is inverse linear, then the implied scaled weights converge pointwise and uniformly in the supremum norm to the true inverse probability weights. 
This implies that the MRI estimator for the treated units $\sum_{i:Z_i=1}w^{{\scriptscriptstyle \text{MRI}}}_i Y^{\text{obs}}_i$ of ${E}\{Y_i(1)\}$ can be viewed as a Horvitz-Thompson IPW estimator $\frac{1}{n}\sum_{i:Z_i=1} Y^{\text{obs}}_i/\hat{e}(\bm{X}_i)$ where $\hat{e}(\bm{X}_i) = (nw^{{\scriptscriptstyle \text{MRI}}}_i)^{-1} = \{n n^{-1}_t + n (\bm{X}_i - \bar{\bm{X}}_t)^\top \bm{S}^{-1}_t (\bar{\bm{X}} - \bar{\bm{X}}_t) \}^{-1}$. 
A similar algebraic equivalence between the IPW estimator and the MRI estimator holds when propensity scores and conditional means are estimated using nonparametric frequency methods (\citealt{hernan2020causal}, Section 13.4). 
Part (b) of Theorem \ref{thm_mri_convergence} provides an analogous result for the MRI weights of the control units. 
However, instead of $e(\bm{x})$, now $1-e(\bm{x})$ needs to be inverse-linear on the covariates. 
Therefore, the linear regression model in the control group implicitly assumes a propensity score model different from the one assumed in the treatment group, since  $e(\bm{x})$ and $1-e(\bm{x})$ cannot be inverse linear simultaneously, unless $e(\bm{x})$ is constant. 
This also means that, unless the propensity score is a constant function of the covariates, the MRI weights for both treated and control units cannot converge simultaneously to their respective true inverse probability weights. 
This condition of constant propensity scores can hold by design in randomized experiments, but is less likely in observational studies. 

We now focus on the convergence of the MRI estimator of the ATE. 
By standard OLS theory, the MRI estimator is consistent for the ATE if both $m_1(\bm{x})$  and $m_0(\bm{x})$ are linear in $\bm{x}$.
The convergence of the MRI weights to the true inverse probability weights in Theorem \ref{thm_mri_convergence} unveils other paths for convergence of the MRI estimator. 
In fact, we obtain five non-nested conditions under which the MRI estimator $\sum_{i:Z_i=1} w^{{\scriptscriptstyle \text{MRI}}}_i Y^{\text{obs}}_i - \sum_{i:Z_i=0} w^{{\scriptscriptstyle \text{MRI}}}_i Y^{\text{obs}}_i$ is consistent for the ATE.

\begin{theorem}
\label{thm_mri_consistency}
The MRI estimator is consistent for the ATE if any of the following conditions holds:
(i) $m_0(\bm{x})$ is linear and $e(\bm{x})$ is inverse linear; (ii) $m_1(\bm{x})$ is linear and $1-e(\bm{x})$ is inverse linear;
(iii) $m_1(\bm{x})$ and $m_0(\bm{x})$ are linear; (iv) $e(\bm{x})$ is constant; (v) $m_1(\bm{x})- m_0(\bm{x})$ is a constant function, $e(\bm{x})$ is linear, and $p^2 \Var(\bm{X}_i\mid Z_i=1) = (1-p)^2 \Var(\bm{X}_i\mid Z_i=0)$ where $p$ is the probability limit of $n_t/n$.

\end{theorem}

Conditions (i), (ii), and (iv) follow from the convergence of the weights described in Theorem \ref{thm_mri_convergence}. Condition (iii) follows from standard OLS theory. Condition (v) relies on an asymptotic equivalence condition between MRI and URI, which we discuss in the Supplementary Material.
Theorem \ref{thm_mri_consistency} shows that the MRI estimator is multiply robust, extending the results of \cite{robins2007comment} and \cite{kline2011oaxaca}. The conditions for multiple robustness are characterized by a combination of conditions on $\{m_1(\cdot), m_0(\cdot), e(\cdot)\}$ jointly, as opposed to conditions on either $\{m_1(\cdot), m_0(\cdot)\}$ or $e(\cdot)$ separately.

While intriguing, this multiple robustness property of the MRI estimator needs to be understood in an adequate context. 
In principle, Theorem \ref{thm_mri_consistency} seems to suggests that many estimators can be doubly robust; the question is under what conditions of the true treatment and outcome models.
For instance, an inverse linear model for $e(\bm{x})$ or $1-e(\bm{x})$ (as in conditions (i) and (ii)) is not very realistic, since the probabilities under an inverse-linear model are not guaranteed to lie inside the (0, 1) range, as noted by \cite{robins2007comment}. Also, even if an inverse-linear model for the treatment is plausible, conditions (i)--(v) may be more stringent in practice than correct specification of either the treatment model or the potential outcome models separately.


Finally, we discuss the asymptotic properties of the URI weights and its associated estimator.
We present these additional results and derivations to the Supplementary Material.
We find that, similar to the MRI weights, the URI weights also converge to the true inverse probability weights, albeit under additional conditions to those in Theorem \ref{thm_mri_convergence}.
A sufficient additional condition for consistency of the URI weights is $p^2 \Var(\bm{X}_i\mid Z_i=1) = (1-p)^2 \Var(\bm{X}_i\mid Z_i=0)$ where $p$ is the probability limit of the proportion of treated units. This condition appears in condition (v) of Theorem \ref{thm_mri_consistency}.
Accordingly, we show that the URI estimator is multiply robust for the ATE, which to our knowledge has not been noted in the literature.
As we show, the required conditions for multiple robustness of URI are even stronger than those for MRI.

\section{Regression diagnostics using the implied weights}
\label{sec_regression}
\subsection{Overview}
The implied weights help us to connect the linear models and observational studies literatures and devise new diagnostics for causal inference using regression.
In this section, we discuss diagnostics based on the implied weights for (i) covariate balance, (ii) model extrapolation, (iii) dispersion and effective sample size, and (iv) influence of a given observation on an estimate of the average treatment effect. We note that the diagnostics in (i), (ii), and (iii) are solely based on the implied weights and do not involve any outcome information. 
In this sense, (i), (ii), and (iii) are part of the design stage of the study \citep{rubin2008objective}. 
In contrast, (iv) requires information from outcomes in addition to the weights, and hence are part of the analysis stage.
We illustrate these weight diagnostics using as a running example the well-known Lalonde study (\citealt{lalonde1986evaluating}) on the impact of a labor training program on earnings. 
The study consists of $n_t = 185$ treated units (enrolled in the program), $n_c = 2490$ control units (not enrolled in program), and $k = 8$ covariates. 
For illustration, here we consider the problem of estimating the ATE.
We implement the implied weights and regression diagnostics in the new \texttt{lmw} package for \texttt{R} available at \texttt{https://github.com/ngreifer/lmw}.

\begin{figure}
\caption{Diagnostics for the URI and MRI weights for the Lalonde observational dataset.}
    \label{fig:diagnostics}
   \centering
    \includegraphics[scale=.9]{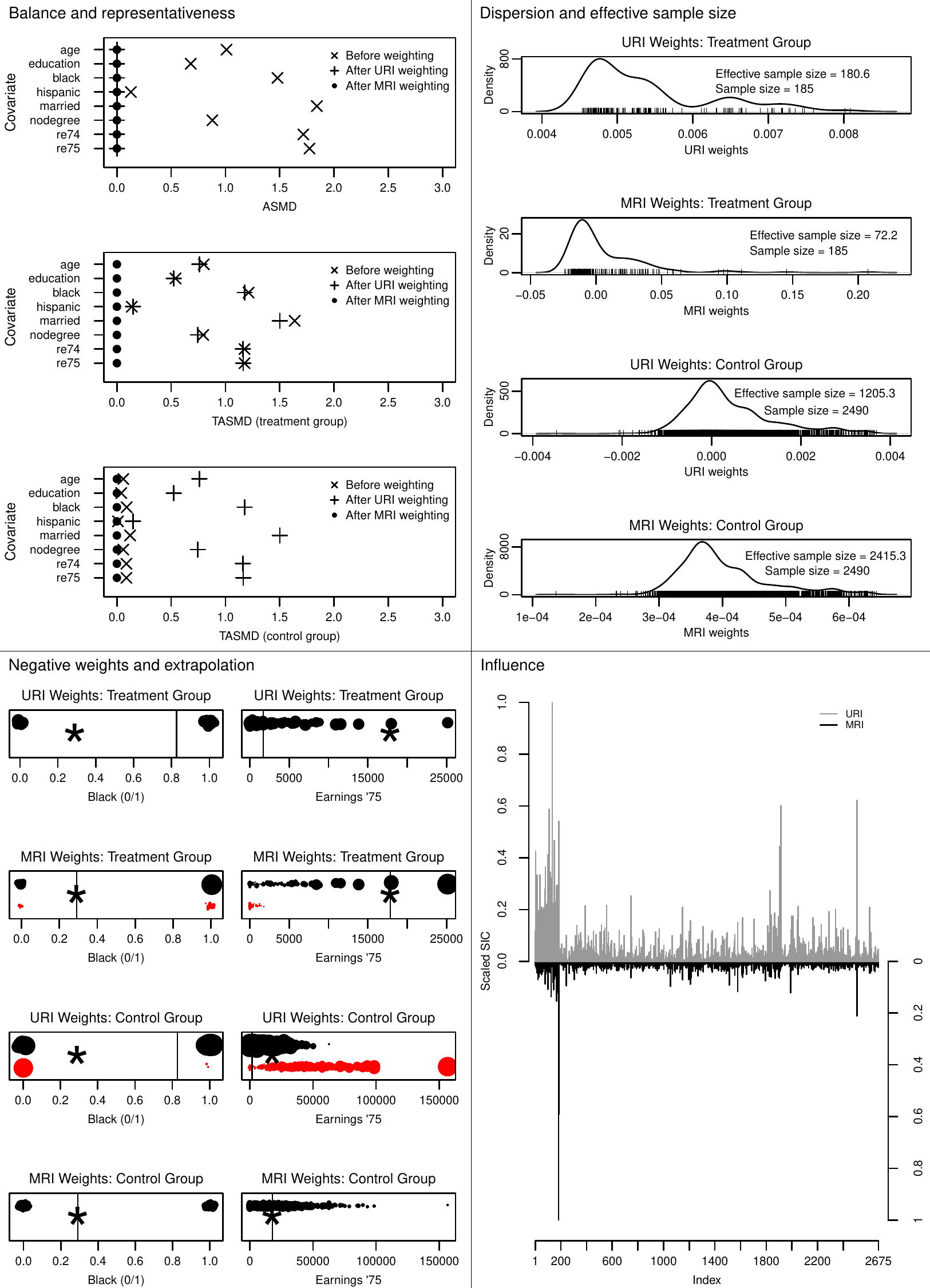}
\end{figure}


\subsection{Covariate balance}
\label{sec_balance_diag}

The implied weights can be used to check balance of the distributions of the covariates in the treatment and control groups relative to a target population. 
As discussed in Section \ref{sec_finite}, although both the URI and MRI weights exactly balance the means of the covariates included in the model, they target different covariate profiles. 
Moreover, neither the URI nor the MRI weights are guaranteed to balance the covariates (or transformations thereof) not included in the model. 
Therefore, it is advisable to check balance on transformations that are not balanced relative to the target by construction. 
A suitable measure for this task is the Target Absolute Standardized Mean Difference (TASMD, \citealt{chattopadhyay2020balancing}), which is defined as the absolute value of the standardized difference between the mean of the covariate transformation in the weighted sample and the corresponding mean in the target population. 
We recommend using the TASMD for balance diagnostics as opposed to the more commonly used Absolute Standardized Mean Difference (ASMD), since it provides a flexible measure of imbalance of a weighted sample relative to arbitrary target profiles, which in principle can represent a single individual. 

The upper left panel of Figure \ref{fig:diagnostics} shows the ASMDs and TASMDs of the eight covariates in the Lalonde study with both URI and MRI. 
The plots illustrate the results in Proposition \ref{prop_finite_sample_properties}. 
In the figure, the first ASMD plot demonstrates the exact mean balancing property of both the URI and the MRI weights, but not relative to a target profile. 
The TASMD plots, on the other hand, provide a more complete picture of the representativeness of URI and the MRI in terms of the first moments of the covariates. 
Since the target profile in this case are the mean of the covariates in the full sample, the MRI weights yield a TASMD of zero for each covariate by construction. 
However, the URI weights do not achieve exact balance relative to the target. 
In fact, in the last TASMD plot we see that the URI weights actually exacerbate the initial imbalances in the control group relative to the target.

\subsection{Extrapolation}
\label{sec_extrapolation}

An important feature of both URI and MRI is that their implied weights can be negative and extreme in magnitude. 
Negative weights are difficult to interpret and, moreover, they can produce estimates that are an extrapolation outside of the support of the available data.
In other words, negative weights can produce H\'{a}jek estimators of average treatment effects that are not sample bounded in the sense of \cite{robins2007comment} (see also \citealt{chattopadhyay2020balancing}). On the other hand, H\'{a}jek estimators with non-negative weights are, by construction, sample bounded.
In some settings, there is no alternative to using negative weights in order to adjust for or balance certain features of the distributions of the observed covariates.
That is, one can only balance the means of such features with negative weights.
If the model behind the adjustments is correctly specified, then extrapolation is not detrimental; however, if the model is misspecified, then it is possible that other features or transformations of the covariates are severely imbalanced and that the estimators are highly biased if these other 
transformations determine $m_1(\cdot)$ and $m_0(\cdot)$.


The bottom left panel of Figure \ref{fig:diagnostics} presents bubble plots of the URI and MRI weights within each treatment group for two covariates, `Black' and `Earnings `75'.
Each bubble represents an observation. 
The size of each bubble is proportional to the absolute value of the corresponding weight. 
A red (respectively, black) bubble indicates a negative (positive) weight. 
The asterisk is the target value of a covariate and the black vertical line represents the weighted average of that covariate in the corresponding treatment group. 
We observe that both MRI and URI produce negative weights. 
Moreover, some of the negative weights are also extreme in magnitude with respect to a particular covariate profile. 
For instance, a control unit with `Earnings `75' greater than $\$$150,000 receives a large negative weight under URI. 
\cite{imbens2015matching} illustrated this phenomenon in the Lalonde study for the ATT using an MRI regression with a single covariate. 
As explained by \cite{imbens2015matching}, OLS linear regression takes linearity very seriously and thus it can render observations with extremely different values from the target profile as highly informative.

\subsection{Weight dispersion and effective sample size}

The variance of the weights is another helpful diagnostic as it directly impacts the variance of the resulting H\'{a}jek estimator under homoscedasticity of the potential outcome models (see, e.g., \citealt{chattopadhyay2020balancing}). However, the variance of the weights often lacks a clear physical interpretation and thus is not directly suitable as a standalone diagnostic in practice. A related but more practical diagnostic is the effective sample size (ESS) of the weighted sample, which provides an intuitive and palpable measure of the number of units that effectively contribute in a weighted sample.

A standard measure of the ESS of a generic weighted sample with non-negative and normalized weights $\{w_1,...,w_{\tilde{n}}\}$ is given by \cite{kish1965survey} as $\tilde{n}_{\text{eff}} = 1/\sum_{i=1}^{\tilde{n}}w^2_i$.
However, for negative weights, this measure may take fractional values or values greater than $\tilde{n}$, which are difficult to interpret. 
To incorporate negative weights, we propose the following modified definition for the ESS: $\tilde{n}_{\text{eff}} = {(\sum_{i=1}^{\tilde{n}}|w_i|)^2}/{\sum_{i=1}^{\tilde{n}}w_i^2}$.
Intuitively, the magnitude of a unit's weight determines its dominance over the other units in the sample. 
Instead of the original weights $w_i$, our definition of the ESS uses Kish's formula on the $|w_i|$. 
The above definition ensures that $\tilde{n}_{\text{eff}} \in [1,\tilde{n}]$. 
When all the weights are non-negative, this definition boils down to Kish's definition of the ESS. 
For both URI and MRI, we recommend computing and reporting the ESS separately for the treatment group and the control group. 

The top right panel of Figure \ref{fig:diagnostics} plots the densities of the URI and MRI weights in each treatment group with their corresponding effective sample sizes. 
While the ESS of URI in the treatment group is almost $98\%$ of the original treatment group size, the ESS of URI in the control group is only $48\%$ of the original control group size. 
This connects to the diagnostics in the previous section where URI proved to have a few units in the control group with extremely large values. 
In contrast, the MRI yields a comparatively high and low ESS in the control and treatment groups, respectively.
\subsection{Influence of a given observation}

Finally, we can characterize the influence of each observation on the regression estimator of the ATE by computing its Sample Influence Curve (SIC,  \citealt{cook1982residuals}). 
In general, consider an estimator $T(\hat{F})$ of a functional $T(F)$, where $F$ is a distribution function and $\hat{F}$ is its corresponding empirical distribution based on a random sample of size $\tilde{n}$. 
The SIC of the $i$th unit in the sample is defined as $\text{SIC}_i = (\tilde{n} - 1) \{T(\hat{F}_{(i)}) - T(\hat{F})\} $, where $\hat{F}_{(i)}$ is the empirical distribution function when the $i$th unit is excluded from the sample. 
$\text{SIC}_i$ is thus proportional to the change in the estimator if the $i$th unit is removed from the data. 
High values of $\text{SIC}_i$ imply a high influence of the $i$th unit on the resulting estimator. 
In Proposition \ref{SIC}, we compute the SIC of the $i$th unit for the URI and MRI estimators of the ATE. 
\begin{proposition} \label{SIC}
For the URI and MRI estimators of the ATE, the Sample Influence Curves of unit $i$ in treatment group $g \in \{t,c\}$ are
$\text{SIC}_i = (n-1)(2Z_i-1)e_iw^{\scriptscriptstyle \textrm{URI}}_i/(1-h_{ii,\bm{D}})$ and $\text{SIC}_i = (n_g-1)e_i w^{\scriptscriptstyle \textrm{MRI}}_i/(1-h_{ii,g})$, respectively, where $e_i$ is the residual for unit $i$ under the corresponding regression model, and $h_{ii}$ and $h_{ii,g}$ are the corresponding leverages of unit $i$.
\end{proposition}
Proposition \ref{SIC} says that the SIC of a unit is a function of its residual, leverage, and implied regression weight. 
A unit can be influential if either its residual, leverage, or weight are large in magnitude. 
In particular, for two units in the same treatment group with the same leverages and residuals, one unit will be more influential than the other if it has a larger weight. However, a large weight alone does not necessarily imply that the unit will have high influence on the corresponding URI or MRI estimator. 
We recommend plotting the absolute values of the SIC in Proposition \ref{SIC} for each observation versus its index as a simple graphical diagnostic of influence. 

In the lower-right panel of Figure \ref{fig:diagnostics}, we plot the absolute-SIC of MRI and URI. 
The values are scaled so that the maximum of the absolute-SIC among the units is one. 
The plot for URI indicates the presence of three highly influential units, whereas, the plot for MRI indicates the presence of one highly influential unit. 
In such cases, one can also plot the SIC versus each covariate to identify which areas of the covariate space lead to these influential units.

\section{Weighted regression and doubly robust estimation}
\label{sec_weighted}


In this section, we extend the results of sections \ref{sec_implied} and \ref{sec_properties} to weighted least squares (WLS) regression. 
In causal inference and sample surveys, WLS can be used to construct doubly robust estimators (e.g., \citealt{kang2007demystifying}). Here we consider extensions of the URI and MRI approaches to WLS, which we call WURI and WMRI respectively. In both WURI and WMRI, a set of base weights $w^\textrm{base}_i$, $i\in \{1,2,...,n\}$, is used in the WLS step to estimate the coefficients of the respective regression models. 
Without loss of generality, we assume that $\sum_{i=1}^{n} w^\textrm{base}_i = 1$ for WURI and that $\sum_{i:Z_i=1} w^\textrm{base}_i = \sum_{i:Z_i=0} w^\textrm{base}_i = 1$ for WMRI. 

In addition to WURI and WMRI, here we analyze the widely used bias-corrected doubly robust (DR) estimator or the augmented inverse probability weighted (AIPW) estimator \citep{robins1994estimation}. 
For the DR estimator, a set of base weights $w^\textrm{base}_i$ with $\sum_{i:Z_i=1} w^\textrm{base}_i = \sum_{i:Z_i=0} w^\textrm{base}_i = 1$ is used in a bias-correction term for the MRI estimator. 
In particular, for $w^\textrm{base}_i$ equal to the inverse probability weights normalized within each treatment group, we can write the DR estimator of the ATE as $\widehat{\text{ATE}}_{\textrm{DR}} = \big[ n^{-1}\sum_{i=1}^{n} \hat{m}_1(\bm{X}_i) + \sum_{i:Z_i=1} w^\textrm{base}_i \{ Y^{\textrm{obs}}_i - \hat{m}_1(\bm{X}_i) \} \big] - \big[ n^{-1}\sum_{i=1}^{n} \hat{m}_0(\bm{X}_i) + \sum_{i:Z_i=0} w^\textrm{base}_i \{ Y^{\textrm{obs}}_i - \hat{m}_0(\bm{X}_i) \} \big]$.

In sections \ref{sec_implied} and \ref{sec_properties}, we showed that the URI and MRI estimators under OLS admit a weighting representation with weights that can be equivalently obtained by solving a quadratic programming problem that minimizes the variance of the weights subject to a normalization constraint and exact mean balancing constraints for the covariates included in the model. 
Here we show that the WURI, WMRI, and DR estimators of the ATE also admit a weighting representation and obtain closed-form expression of their implied weights.


\begin{theorem} 
\label{thm_general}
Consider the following quadratic programming problem in the control group
\begin{equation*}\label{calibration}
\begin{aligned} 
& \underset{\bm{w}}{\text{minimize}}
& & \sum_{i: Z_i=0}\frac{(w_i - \tilde{w}^\textrm{base}_i)^2}{w^\textrm{scale}_i}\\
& \text{subject to}
& & \mid\sum_{i: Z_i=0} w_i \bm{X}_i - \bm{X}^*\mid \leq \bm{\delta} \\
& & & \sum_{i: Z_i=0} w_i=1\\
\end{aligned}
\end{equation*}
where $\tilde{w}^\textrm{base}_i$ are normalized base weights in the control group, $w^\textrm{scale}_i$ are scaling weights, and $\bm{X}^* \in \mathbb{R}^k$ is a covariate profile, all of them determined by the investigator.
Then, for $\bm{\delta} = \bm{0}$ the solution to this problem is
$$
w_i = \tilde{w}^\textrm{base}_i + w^\textrm{scale}_i(\bm{X}_i - \bar{\bm{X}}_c^\textrm{scale})^\top ( \bm{S}_c^\textrm{scale}/n_c)^{-1} (\bm{X}^* - \bar{\bm{X}}_c^\textrm{base}),
$$ 
where $\bar{\bm{X}}_c^\textrm{scale} = (\sum_{i:Z_i=0} w^\textrm{scale}_i \bm{X}_i)/(\sum_{i:Z_i=0} w^\textrm{scale}_i) $, $\bar{\bm{X}}_c^\textrm{base} = \sum_{i:Z_i=0} \tilde{w}^\textrm{base}_i \bm{X}_i$, and $\bm{S}_c^\textrm{scale} = n_c \sum_{i:Z_i=0} w^\textrm{scale}_i (\bm{X}_i - \bar{\bm{X}}_c^\textrm{scale}) (\bm{X}_i - \bar{\bm{X}}_c^\textrm{scale})^\top $.
Further, as special cases the implied weights of the WURI, WMRI, and DR estimators for the ATE are
\begin{enumerate}[label=(\alph*)]
    \item WURI: $\tilde{w}^\textrm{base}_i = w^\textrm{base}_i/(\sum_{j: Z_{j} = 0}w^\textrm{base}_j)$, $w^\textrm{scale}_i = w^\textrm{base}_i$, $\bm{X}^* = n^{-1}_c\bm{S}_c^\textrm{scale} (n^{-1}_t\bm{S}_t^\textrm{scale} + n^{-1}_c\bm{S}_c^\textrm{scale})^{-1} \bar{\bm{X}}_t^\textrm{scale} + n^{-1}_t \bm{S}_t^\textrm{scale} \left(n^{-1}_t\bm{S}_t^\textrm{scale} + n^{-1}_c \bm{S}_c^\textrm{scale}\right)^{-1} \bar{\bm{X}}_c^\textrm{scale}$. 
    
    \item WMRI: $\tilde{w}^\textrm{base}_i = w^\textrm{scale}_i = w^\textrm{base}_i$, $\bm{X}^* = \bar{\bm{X}}$.
    
    \item DR: $\tilde{w}^\textrm{base}_i = w^\textrm{base}_i = \{1-\hat{e}(\bm{X}_i)\}^{-1}/\sum_{j:Z_j=0}\{1-\hat{e}(\bm{X}_j)\}^{-1}$, $w^\textrm{scale}_i = 1$, $\bm{X}^* = \bar{\bm{X}}$.
\end{enumerate}
Here, $\bar{\bm{X}}_t^\textrm{scale} = (\sum_{i:Z_i=1} w^\textrm{scale}_i \bm{X}_i)/(\sum_{i:Z_i=1} w^\textrm{scale}_i)$ and $\bm{S}_t^\textrm{scale} = n_t \sum_{i:Z_i=1} w^\textrm{scale}_i (\bm{X}_i - \bar{\bm{X}}_t^\textrm{scale}) (\bm{X}_i - \bar{\bm{X}}_t^\textrm{scale})^\top$.
The weights for the treated units are obtained analogously. \end{theorem}

Theorem \ref{thm_general} shows that the implied weights of WURI, WMRI, and DR estimators can all be obtained by solving one quadratic programming problem. 
Coming back to question about the connection between experiments and regression, this theorem also characterizes the covariate adjustments of WURI, WMRI, and DR estimators in terms of covariate balance and representativeness, weight dispersion, and sample boundedness.
First, the implied weights of all the estimators exactly balance the means of the covariates, but towards different target profiles. While both WMRI and DR estimators balance covariates towards the overall sample mean, WURI balances them towards a target that is harder to interpret. 
Second, the weights minimize a general measure of weight dispersion, e.g., the WURI and WMRI weights minimize a Chi-square-type distance from the base weights, whereas the DR weights minimize the Euclidean distance from the base weights. 
For uniform base weights, these measures of dispersion boil down to the variance of the weights. 
Finally, the weights can be negative and hence can extrapolate, since the underlying quadratic programs do not impose any non-negativity constraint on the weights.

In terms of the above features, the quadratic programming problem in Theorem \ref{thm_general} also clarifies similarities and differences of the WURI, WMRI, and DR weights with the stable balancing weights (\citealt{zubizarreta2015stable}). 
The stable balancing weights are the weights of minimum variance that sum to one, are non-negative, and approximately balance the means of functions of the covariates towards a pre-specified profile. 
Here, the non-negativity constraints are used to produce a sample bounded estimator and the approximate balance constraints help to trade bias for variance. 
Methodologically, imposing these two types of constraints may correspond to fitting a new type of regularized and sample-bounded regression. 


\section{Conclusion}
\label{sec_conclusion}

Across the health and social sciences, linear regression is extensively used in observational studies to adjust for covariates and estimate the  effects of treatments. 
In this paper, we demonstrated how linear regression adjustments in observational studies emulate key features of randomized experiments.
For this, we represented regression estimators as weighting estimators and derived their implied weights.
We obtained new closed-form, finite-sample expressions for the weights for various types of estimators (URI, MRI, AIPW) based on multivariate linear regression models under both ordinary and weighted least squares. 
Among others, we obtained a closed-form expression of the covariate profile targeted by the implied weights, which, in turn, characterizes the implied target population of regression. 
In this regard, we showed that URI estimators may distort the structure of the study sample in such a way that the resulting estimator is biased for the average treatment effect in the target population.
In contrast, MRI estimators preserve the structure of this population in terms of its first moments, albeit with negative weights. 

This implied weighting framework allowed us to connect ideas from the regression modeling and the causal inference literature, and leveraging this connection, to propose a set of new regression diagnostics for causal inference. 
These diagnostics allow researchers to (i) evaluate the degree of covariate balance of each weighted treatment group relative to the target population, (ii) compute the effective sample size of each treatment group after regression adjustments, (iii) assess the degree of extrapolation and sample-boundedness of the regression estimates, and (iv) measure the influence of a given observation on an average treatment effect estimate. 
We also showed that the URI and MRI estimators are both multiply robust for the average treatment effect, albeit under conditions that impose stringent functional forms on the treatment model, the potential outcome models, or a combination of both. 

Extensions of this work include the implied weights of regression with continuous and multi-valued treatments, and regression after matching. 
As future work, we plan to extend the implied weights framework to analyze instrumental variables methods \citep{abadie2003semiparametric} and fixed-effects regressions in difference-in-differences settings \citep{ding2019bracketing}. 

\bibliographystyle{chicago}
\bibliography{bib2022}

\pagebreak
\section{Supplementary materials}

\section{Additional theoretical results}

\subsection{Derivation of the WURI weights}
\label{wuri_derive}
Here we use the notations of Theorem 3. In WURI, we fit the model $\bm{y} =  \mu \bm{1} + \underline{\bm{X}}\bm{\beta} + \tau \bm{Z} + \bm{\epsilon} $ using WLS with the base weights $(w^{\text{base}}_1,...,w^{\text{base}}_n)$. Without loss of generality, assume $\sum_{i=1}^{n}w^{\text{base}}_i =1$. Also, for all $i \in \{1,...,n\}$, let $w^{\text{scale}}_i = w^{\text{base}}_i$. Denote the design matrix based on the full sample, treatment group, and control group as $\underline{\tilde{\bm{X}}}$, $\underline{\tilde{\bm{X}}}_t$, and $\underline{\tilde{\bm{X}}}_c$ respectively. Also, let $\bm{W}^{\frac{1}{2}} = diag((w^{\text{base}}_1)^{0.5},...,(w^{\text{base}}_n)^{0.5})$, $\bm{W} = \bm{W}^{\frac{1}{2}}\bm{W}^{\frac{1}{2}}$,  $\dbar{\bm{X}} = \bm{W}^{\frac{1}{2}} \underline{\tilde{\bm{X}}} = \bm{W}^{\frac{1}{2}}
(\bm{1} , \bm{X})$, $\dbar{\bm{y}} = \bm{W}^{\frac{1}{2}} \bm{y}$, $\dbar{\bm{Z}} = \bm{W}^{\frac{1}{2}}\bm{Z}$. The objective function under WLS is given by
\begin{align} 
\argmin_{\mu,\bm{\beta}, \tau} \Big(\bm{y} - \underline{\tilde{\bm{X}}}\begin{psmallmatrix}
\mu\\
\bm{\beta}
\end{psmallmatrix} - \tau \bm{Z} \Big)^\top \bm{W} \Big(\bm{y} - \underline{\tilde{\bm{X}}}\begin{psmallmatrix}
\mu\\
\bm{\beta}
\end{psmallmatrix} - \tau \bm{Z} \Big) \nonumber\\
= \argmin_{\mu,\bm{\beta}, \tau} \Big(\dbar{\bm{y}} - \dbar{\bm{X}}\begin{psmallmatrix}
\mu\\
\bm{\beta}
\end{psmallmatrix} - \tau \dbar{\bm{Z}} \Big)^\top \Big(\dbar{\bm{y}} - \dbar{\bm{X}}\begin{psmallmatrix}
\mu\\
\bm{\beta}
\end{psmallmatrix} - \tau \dbar{\bm{Z}} \Big).
\label{s1.1.1}
\end{align}
Therefore, the objective function under WLS is equivalent to that under OLS for a linear regression of $\dbar{\bm{y}}$ on $\dbar{\bm{X}}$ and $\dbar{\bm{Z}}$. Let $\underline{\bm{I}}$ be the identity matrix of order $k+1$ and $\mathcal{P}_{\dbar{\bm{X}}}$ be the projection matrix onto the column space of $\dbar{\bm{X}}$. Using the Frisch-Waugh-Lovell Theorem (\citealt{frisch1933partial, lovell1963seasonal}), we can write the corresponding estimator of $\tau$ as
\begin{equation}
\hat{\tau} = (\dbar{\bm{Z}}^\top (\underline{\bm{I}} - \mathcal{P}_{\dbar{\bm{X}}})\dbar{\bm{y}})/(\dbar{\bm{Z}}^\top (\underline{\bm{I}} - \mathcal{P}_{\dbar{\bm{X}}})\dbar{\bm{Z}}) = \bm{l}^\top \bm{y},
\label{s1.1.2}
\end{equation}
where $\bm{l} = (\bm{W}^{\frac{1}{2}}(I - \mathcal{P}_{\dbar{\bm{X}}})\dbar{\bm{Z}})/(\dbar{\bm{Z}}^\top (I - \mathcal{P}_{\dbar{\bm{X}}})\dbar{\bm{Z}})$. Denote $m_t = \sum_{i:Z_i=1}w^{\text{base}}_i $ and $m_c = \sum_{i:Z_i=0}w^{\text{base}}_i $. By assumption, $m_t + m_c = 1$.
Now, the denominator $\dbar{\bm{Z}}^\top (\underline{\bm{I}} - \mathcal{P}_{\dbar{\bm{X}}})\dbar{\bm{Z}} = m_t - \bm{Z}^\top \bm{W} \underline{\tilde{\bm{X}}}(\underline{\tilde{\bm{X}}}^\top \bm{W}\underline{\tilde{\bm{X}}})^{-1} \underline{\tilde{\bm{X}}}^\top \bm{W} \bm{Z} = m_t \big\{1 - m_t (1, \bar{\bm{X}}^{\text{scale}\top}_{t})(\underline{\tilde{\bm{X}}}^\top \bm{W}\underline{\tilde{\bm{X}}})^{-1} (1, \bar{\bm{X}}^{\text{scale}\top}_{t})^\top \big\} $. Similarly, the numerator is
\begin{equation}
\bm{W}^{\frac{1}{2}}(\underline{\bm{I}} - \mathcal{P}_{\dbar{\bm{X}}})\dbar{\bm{Z}} = \bm{WZ} - \bm{W}\underline{\tilde{\bm{X}}}(\underline{\tilde{\bm{X}}}^\top \bm{W} \underline{\tilde{\bm{X}}})^{-1} \underline{\tilde{\bm{X}}}^{\top} \bm{W} \bm{Z}    
\label{s1.1.3}
\end{equation}
To simplify the expression, let us assume, without loss of generality, that the first $n_t$ units in the sample are in the treatment group and the rest are in the control group. Moreover, let $\bm{w}_t = (w^{\text{base}}_1,....,w^{\text{base}}_{n_t})$ and $\bm{w}_c = (w^{\text{base}}_{n_t+1},....,w^{\text{base}}_{n})$ be the vector of base weights for the treatment and control group respectively. This implies
\begin{align}
\bm{W}^{\frac{1}{2}}(I - \mathcal{P}_{\dbar{\bm{X}}})\dbar{\bm{Z}} & = \begin{psmallmatrix}
\bm{w}_t\\
\bm{0}
\end{psmallmatrix} - \bm{W} \begin{psmallmatrix}
\underline{\tilde{\bm{X}}}_t (\underline{\tilde{\bm{X}}}^\top \bm{W} \underline{\tilde{\bm{X}}})^{-1} \underline{\tilde{\bm{X}}}^\top_t \bm{w}_t\\
\underline{\tilde{\bm{X}}}_c (\underline{\tilde{\bm{X}}}^\top \bm{W} \underline{\tilde{\bm{X}}})^{-1} \underline{\tilde{\bm{X}}}^\top_t \bm{w}_t 
\end{psmallmatrix}
\label{s1.1.4}
\end{align}
From Equation \ref{s1.1.2} we get that we can write $\hat{\tau}$ as $\hat{\tau} = \sum_{i:Z_i=1}w_i Y^{\text{obs}}_i  - \sum_{i:Z_i=0}w_i Y^{\text{obs}}_i $. Let $\bar{\bm{X}}^{\text{scale}} = \sum_{i=1}^{n} w^{\text{scale}}_i\bm{X}_i$.
From Equation \ref{s1.1.4}, it follows that if the $i$th unit belongs to the treatment group then

\begin{align}
w_i &= \frac{w^{\text{base}}_i \big\{ 1 - (1,\bm{X}^\top_i) (\underline{\tilde{\bm{X}}}^\top \bm{W} \underline{\tilde{\bm{X}}})^{-1} \underline{\tilde{\bm{X}}}_t^\top \bm{w}_t  \big\}}{  m_t \big(1 - m_t (1, \bar{\bm{X}}^{\text{scale}\top}_{t})(\underline{\tilde{\bm{X}}}^\top \bm{W}\underline{\tilde{\bm{X}}})^{-1} (1, \bar{\bm{X}}^{\text{scale}\top}_{t})^\top \big)} \nonumber\\
& =  \frac{w^{\text{base}}_i \big(1 - m_t(0,(\bm{X}_i - \bar{\bm{X}}^{\text{scale}}_{t})^\top) (\underline{\tilde{\bm{X}}}^\top \bm{W} \underline{\tilde{\bm{X}}})^{-1} (1, \bar{\bm{X}}^{\text{scale}\top}_{t})^\top  -  m_t (1,\bar{\bm{X}}^{\text{scale}\top}_{t})  (\underline{\tilde{\bm{X}}}^\top \bm{W} \underline{\tilde{\bm{X}}})^{-1} (1,\bar{\bm{X}}^{\text{scale}\top}_{t})^\top  \big)}{ m_t \big(1 - m_t (1, \bar{\bm{X}}^{\text{scale}\top}_{t})(\underline{\tilde{\bm{X}}}^\top \bm{W}\underline{\tilde{\bm{X}}})^{-1} (1, \bar{\bm{X}}^{\text{scale}\top}_{t})^\top \big)} \nonumber\\
& = w^{\text{base}}_i \Big[ \frac{1}{m_t} + \frac{(0,(\bm{X}_i - \bar{\bm{X}}^{\text{scale}}_{t})^\top) (\underline{\tilde{\bm{X}}}^\top \bm{W} \underline{\tilde{\bm{X}}})^{-1} (0,(\bar{\bm{X}}^{\text{scale}} - \bar{\bm{X}}^{\text{scale}}_{t})^\top) }{1 - m_t (1, \bar{\bm{X}}^{\text{scale}\top}_{t})(\underline{\tilde{\bm{X}}}^\top \bm{W}\underline{\tilde{\bm{X}}})^{-1} (1, \bar{\bm{X}}^{\text{scale}\top}_{t})^\top }   \Big] \nonumber\\
& = w^{\text{base}}_i \Big[ \frac{1}{m_t} + \frac{(0,(\bm{X}_i - \bar{\bm{X}}^{\text{scale}}_{t})^\top) (\underline{\tilde{\bm{X}}}^\top \bm{W} \underline{\tilde{\bm{X}}})^{-1} (0,(\bar{\bm{X}}^{\text{scale}} - \bar{\bm{X}}^{\text{scale}}_{t})^\top) }{1 - m_t \big( 1 + (0, (\bar{\bm{X}}^{\text{scale}}-\bar{\bm{X}}^{\text{scale}}_{t})^\top)(\underline{\tilde{\bm{X}}}^\top \bm{W}\underline{\tilde{\bm{X}}})^{-1} (0, (\bar{\bm{X}}^{\text{scale}}-\bar{\bm{X}}^{\text{scale}}_{t})^\top)^\top \big)}   \Big] \nonumber\\
& = w^{\text{base}}_i \Big[ \frac{1}{m_t} + \frac{(\bm{X}_i - \bar{\bm{X}}^{\text{scale}}_{t})^\top (\frac{\bm{S}^{\text{scale}}}{n})^{-1} (\bar{\bm{X}}^{\text{scale}} - \bar{\bm{X}}^{\text{scale}}_{t}) }{1 - m_t \big( 1 + (\bar{\bm{X}}^{\text{scale}}-\bar{\bm{X}}^{\text{scale}}_{t})^\top(\frac{\bm{S}^{\text{scale}}}{n})^{-1} (\bar{\bm{X}}^{\text{scale}}-\bar{\bm{X}}^{\text{scale}}_{t})  \big)}  \Big],
\label{s1.1.5}
\end{align}
where $\bm{S}^{\text{scale}} = n\big(\sum_{j=1}^{n} w^{\text{scale}}_{j} \bm{X}_{j} \bm{X}^\top_{j} -  \bar{\bm{X}}^{\text{scale}} \bar{\bm{X}}^{\text{scale}\top}\big) $. The second equality holds since $\underline{\tilde{\bm{X}}}_t^\top \bm{w}_t = m_t(1, \bar{\bm{X}}^{\text{scale}\top}_{t})^\top$. The third and fourth equality hold since $(1, \bar{\bm{X}}^{\text{scale}\top})^\top = \underline{\tilde{\bm{X}}}^\top \bm{W} \underline{\tilde{\bm{X}}} \bm{e}_1$, where $\bm{e}_1 = (1,0,0,...,0)$ is the first standard basis vector of $\mathbb{R}^{k+1}$. To see the fifth equality, we first see that $\underline{\tilde{\bm{X}}}^\top \bm{W} \underline{\tilde{\bm{X}}} = \begin{pmatrix}
    1 & \bar{\bm{X}}^{\text{scale}\top} \\
    \bar{\bm{X}}^{\text{scale}} & \underline{\bm{X}}^\top \bm{W} \underline{\bm{X}}
\end{pmatrix}$. Let $(\underline{\tilde{\bm{X}}}^\top \bm{W} \underline{\tilde{\bm{X}}})^{-1} = \begin{pmatrix}
    {B_{11}}^{(1\times 1)} & {B_{12}}^{(1\times k))} \\
    {B_{21}}^{(k \times 1)} & {B_{22}}^{(k \times k)}
\end{pmatrix}$. Using the formula for the inverse of a partitioned matrix, we get $B_{22} = \big[\underline{\bm{X}}^\top \bm{W} \underline{\bm{X}} - \bar{\bm{X}}^{\text{scale}} \bar{\bm{X}}^{\text{scale}\top} \big]^{-1} = \big[ \sum_{j=1}^{n} w^{\text{scale}}_{j} \bm{X}_{j} \bm{X}^\top_{j} -  \bar{\bm{X}}^{\text{scale}} \bar{\bm{X}}^{\text{scale}\top} \big]^{-1} = (\frac{\bm{S}^{\text{scale}}}{n})^{-1} $. Now, we observe that $\bm{S}^{\text{scale}} = n(\frac{\bm{S}^{\text{scale}}_t}{n_t} + \frac{\bm{S}^{\text{scale}}_c}{n_c} + m_t m_c (\bar{\bm{X}}^{\text{scale}}_{t} - \bar{\bm{X}}^{\text{scale}}_c)(\bar{\bm{X}}^{\text{scale}}_{t} - \bar{\bm{X}}^{\text{scale}}_c)^\top )$. This implies,
\begin{align}
&\big(\frac{\bm{S}^{\text{scale}}_t}{n_t} + \frac{\bm{S}^{\text{scale}}_c}{n_c} \big)\big(\frac{\bm{S}^{\text{scale}}}{n}\big)^{-1} (\bar{\bm{X}}^{\text{scale}}_{t} - \bar{\bm{X}}^{\text{scale}}_c) \nonumber\\ &= \big(\frac{\bm{S}^{\text{scale}}}{n} - m_t m_c (\bar{\bm{X}}^{\text{scale}}_{t} - \bar{\bm{X}}^{\text{scale}}_c)(\bar{\bm{X}}^{\text{scale}}_{t} - \bar{\bm{X}}^{\text{scale}}_c)^\top  \big)\big(\frac{\bm{S}^{\text{scale}}}{n}\big)^{-1} (\bar{\bm{X}}^{\text{scale}}_{t} - \bar{\bm{X}}^{\text{scale}}_c) \nonumber\\
&= (\bar{\bm{X}}^{\text{scale}}_{t} - \bar{\bm{X}}^{\text{scale}}_c) - \chi (\bar{\bm{X}}^{\text{scale}}_{t} - \bar{\bm{X}}^{\text{scale}}_c) \nonumber\\
&\implies \big(\frac{\bm{S}^{\text{scale}}_t}{n_t} + \frac{\bm{S}^{\text{scale}}_c}{n_c} \big)^{-1} (\bar{\bm{X}}^{\text{scale}}_{t} - \bar{\bm{X}}^{\text{scale}}_c) = \big(\frac{\bm{S}^{\text{scale}}}{n}\big)^{-1} (\bar{\bm{X}}^{\text{scale}}_{t} - \bar{\bm{X}}^{\text{scale}}_c)/(1-\chi) ,
\label{s1.1.6}
\end{align}  
where $\chi = m_t m_c (\bar{\bm{X}}^{\text{scale}}_{t} - \bar{\bm{X}}^{\text{scale}}_c)^\top \big(\frac{\bm{S}^{\text{scale}}}{n}\big)^{-1} (\bar{\bm{X}}^{\text{scale}}_{t} - \bar{\bm{X}}^{\text{scale}}_c)$. From Equations \ref{s1.1.5} and \ref{s1.1.6}, we get
\begin{align}
w_i & = w^{\text{base}}_i \Big[ \frac{1}{m_t} + \frac{(\bm{X}_i - \bar{\bm{X}}^{\text{scale}}_{t})^\top (\frac{\bm{S}^{\text{scale}}}{n})^{-1} (\bar{\bm{X}}^{\text{scale}} - \bar{\bm{X}}^{\text{scale}}_{t}) }{m_c (1-\chi)}  \Big] \nonumber\\
&= w^{\text{base}}_i \Big[ \frac{1}{m_t} + \frac{1}{m_c}(\bm{X}_i - \bar{\bm{X}}^{\text{scale}}_{t})^\top \big(\frac{\bm{S}^{\text{scale}}_t}{n_t} + \frac{\bm{S}^{\text{scale}}_c}{n_c} \big)^{-1} (\bar{\bm{X}}^{\text{scale}} - \bar{\bm{X}}^{\text{scale}}_{t}) \Big] \nonumber\\
&= \tilde{w}^{\text{base}}_i + \frac{w^{\text{scale}}_i}{m_c}(\bm{X}_i - \bar{\bm{X}}^{\text{scale}}_{t})^\top \big(\frac{\bm{S}^{\text{scale}}_t}{n_t} + \frac{\bm{S}^{\text{scale}}_c}{n_c} \big)^{-1} (\bar{\bm{X}}^{\text{scale}} - \bar{\bm{X}}^{\text{scale}}_{t}),
\label{s1.1.7}
\end{align}
where in the last equality, we have used the fact that the base weights are same as the scaling weights. Now, let $\bm{X}^* = \frac{\bm{S}_c^\text{scale}}{n_c} \left(\frac{\bm{S}_t^\text{scale}}{n_t} + \frac{\bm{S}_c^\text{scale}}{n_c}\right)^{-1} \bar{\bm{X}}_t^\text{scale} + \frac{\bm{S}_t^\text{scale}}{n_t} \left(\frac{\bm{S}_t^\text{scale}}{n_t} + \frac{\bm{S}_c^\text{scale}}{n_c}\right)^{-1} \bar{\bm{X}}_c^\text{scale}$. By simple substitution and using the fact that $\bar{\bm{X}}^{\text{base}} = \bar{\bm{X}}^{\text{scale}}$, we get
\begin{equation}
    \Big(\frac{\bm{S}^{\text{scale}}_t}{n_t} \Big)^{-1} (\bm{X}^* - \bar{\bm{X}}^{\text{base}}_t) = \frac{1}{m_c} \big(\frac{\bm{S}^{\text{scale}}_t}{n_t} + \frac{\bm{S}^{\text{scale}}_c}{n_c} \big)^{-1} (\bar{\bm{X}}^{\text{scale}} - \bar{\bm{X}}^{\text{scale}}_{t}).
    \label{s1.1.8}
\end{equation}
This implies,
\begin{equation}
    w_i = \tilde{w}^\text{base}_i + w^\text{scale}_i(\bm{X}_i - \bar{\bm{X}}_t^\text{scale})^\top \left( \frac{\bm{S}_t^\text{scale}}{n_t} \right)^{-1} (\bm{X}^* - \bar{\bm{X}}_t^\text{base}).
\end{equation}
Using the structural symmetry between the treatment and control group, it follows that, if the $i$th unit belongs to the control group, then
\begin{equation*}
w_i = \tilde{w}^\text{base}_i + w^\text{scale}_i(\bm{X}_i - \bar{\bm{X}}_c^\text{scale})^\top \left( \frac{\bm{S}_c^\text{scale}}{n_c} \right)^{-1} (\bm{X}^* - \bar{\bm{X}}_c^\text{base}).
\end{equation*}

\subsection{Derivation of WMRI weights} 
\label{wmri_derive}
Here we use the notations of Theorem 3. Without loss of generality, let us assume that the first $n_c$ units in the sample belong to the control group. In WMRI, we fit the linear model $\bm{y}_c = \beta_{0c} \bm{1} + \underline{\bm{X}}_c \bm{\beta}_{1c} + \bm{\epsilon}_c$ in the control group using WLS with the base weights $(w^{\text{base}}_1,...,w^{\text{base}}_{n_c})$, where $\sum_{i=1}^{n_c}w^{\text{base}}_i = 1$. For all $i\in \{1,...,n_c\}$, let $w^{\text{scale}}_i = w^{\text{base}}_i$. Also, let $\bm{W}_c = diag(w^{\text{base}}_1,...,w^{\text{base}}_{n_c})$.
The WLS estimator of the parameter vector $\bm{\beta}_c = (\beta_{0c}, \bm{\beta}^\top_{1c})^\top$ is given by $\hat{\bm{\beta}}_{c} = (\underline{\tilde{\bm{X}}}^\top \bm{W}_c \underline{\tilde{\bm{X}}})^{-1} \underline{\tilde{\bm{X}}}^\top \bm{W}_c \bm{y}_c $. The estimated mean function of the control potential outcome for any unit with a generic covariate profile $\bm{x}$ is $\hat{m}_0(\bm{x}) = \hat{\beta}_{0c} + \hat{\bm{\beta}}^\top_{1c} \bm{x} = (\bm{w}_c)^\top \bm{y}_c$, where $\bm{w}_c = (w_1,...,w_{n_c})^\top$ is given by
\begin{align}
\bm{w}_c &= \bm{W}_c^\top \underline{\tilde{\bm{X}}}_c  (\underline{\tilde{\bm{X}}}^\top_c \bm{W}_c \underline{\tilde{\bm{X}}}_c)^{-1} (1, \bm{x}^{\top})^\top \nonumber \\
&= \bm{W}_c \underline{\tilde{\bm{X}}}_c  (\underline{\tilde{\bm{X}}}^\top_c \bm{W}_c \underline{\tilde{\bm{X}}}_c)^{-1} (1, \bar{\bm{X}}^{\text{scale}\top}_c)^\top + \bm{W}_c \underline{\tilde{\bm{X}}}_c  (\underline{\tilde{\bm{X}}}^\top_c \bm{W}_c \underline{\tilde{\bm{X}}}_c)^{-1} (0, (\bm{x} - \bar{\bm{X}}^{\text{scale}}_c)^\top)^\top \nonumber\\ 
&= \bm{W}_c \bm{1} + \bm{W}_c \underline{\tilde{\bm{X}}}_c  (\underline{\tilde{\bm{X}}}^\top_c \bm{W}_c \underline{\tilde{\bm{X}}}_c)^{-1} (0, (\bm{x} - \bar{\bm{X}}^{\text{scale}}_c)^\top)^\top
\label{s1.2.1}
\end{align}
The second inequality is obtained by noting that $(1, \bar{\bm{X}}^{\text{scale}\top}_c)^\top = \underline{\tilde{\bm{X}}}^\top_c \bm{W}_c \bm{1}  = \underline{\tilde{\bm{X}}}^\top_c \bm{W}_c \underline{\tilde{\bm{X}}}_c \bm{e}_1$, where $\bm{e}_1 = (1,0,...,0)^\top$ is the first standard basis vector of $\mathbb{R}^{k+1}$. Therefore, $\hat{m}_0(\bm{x}) = \sum_{i:Z_i=0} w_i Y^{\text{obs}}_i$, where  
\begin{align}
w_i &= w^{\text{base}}_i \big\{ 1 + (1,\bm{X}^\top_i) (\underline{\tilde{\bm{X}}}^\top_c \bm{W}_c \underline{\tilde{\bm{X}}}_c)^{-1} (0, (\bm{x} - \bar{\bm{X}}^{\text{scale}}_c)^\top)^\top   \big\} \nonumber\\
&= w^{\text{base}}_i \big\{ 1 + (0,(\bm{X}_i - \bar{\bm{X}}^{\text{scale}}_c)^\top) (\underline{\tilde{\bm{X}}}^\top_c \bm{W}_c \underline{\tilde{\bm{X}}}_c)^{-1} (0, (\bm{x} - \bar{\bm{X}}^{\text{scale}}_c)^\top)^\top   \big\} \nonumber\\
&= w^{\text{base}}_i  + w^{\text{base}}_i(\bm{X}_i - \bar{\bm{X}}^{\text{scale}}_c)^\top \big(\frac{\bm{S}^{\text{scale}}_c}{n_c}\big)^{-1} (\bm{x} - \bar{\bm{X}}^{\text{scale}}_c).
\label{s1.2.2}
\end{align}
The last equality holds by similar arguments as in the WURI case. Using the fact that the base weights and the scaling weights are the same and that the base weights are normalized, we get
\begin{equation}
    w_i = \tilde{w}^{\text{base}}_i  + w^{\text{scale}}_i(\bm{X}_i - \bar{\bm{X}}^{\text{scale}}_c)^\top \big(\frac{\bm{S}^{\text{scale}}_c}{n_c}\big)^{-1} (\bm{x} - \bar{\bm{X}}^{\text{base}}_c).
    \label{s1.2.3}
\end{equation}
Similarly, if we fit the linear model $\bm{y}_t = \beta_{0t} \bm{1} + \bm{X}_c \bm{\beta}_{1t} + \bm{\epsilon}_c $ in the treatment group using WLS then, by similar steps, the estimated mean function of the treatment potential outcome for any unit with covariate profile $\bm{x}^*$ is given by $\hat{m}_1(\bm{x}^*) = \hat{\beta}_{0t} + \hat{\bm{\beta}}^\top_{1t} \bm{x}^* = \sum_{i=1}^{n_t} w_i Y^{\text{obs}}_{i,t} $, where 
\begin{equation}
    w_i = \tilde{w}^{\text{base}}_i  + w^{\text{scale}}_i(\bm{X}_i - \bar{\bm{X}}^{\text{scale}}_t)^\top \big(\frac{\bm{S}^{\text{scale}}_t}{n_t}\big)^{-1} (\bm{x} - \bar{\bm{X}}^{\text{base}}_t).
    \label{s1.2.4}
\end{equation}
Therefore, we have the following results.
\begin{enumerate}
    \item By linearity, he WMRI estimator of the ATE is $\widehat{\text{ATE}} = \hat{m}_1(\bar{\bm{X}}) - \hat{m}_0(\bar{\bm{X}}) = \sum_{i:Z_i=1} w_i Y^{\text{obs}}_i - \sum_{i:Z_i=0} w_i Y^{\text{obs}}_i$, where $w_i$ has the form as given in Equations \ref{s1.2.3} and \ref{s1.2.4} with $\bm{x}$ replaced by $\bar{\bm{X}}$.
    \item By linearity, he WMRI estimator of the ATT is $\widehat{\text{ATT}} = \bar{Y}_t - \hat{m}_0(\bar{\bm{X}}_t) = \bar{Y}_t - \sum_{i:Z_i=0} w_i Y^{\text{obs}}_i$, where $w_i$ has the form as given in Equation \ref{s1.2.3} with $\bm{x}$ replaced by $\bar{\bm{X}}_t$.
    \item The WMRI estimator of the CATE at covariate profile $\bm{x}^*$ is given by $\widehat{\text{CATE}}(\bm{x}^*) = \hat{m}_1(\bm{x}^*) - \hat{m}_0(\bm{x}^*) = \sum_{i:Z_i=1} w_i Y^{\text{obs}}_i - \sum_{i:Z_i=0} w_i Y^{\text{obs}}_i$, where $w_i$ has the form as given in Equations \ref{s1.2.3} and \ref{s1.2.4} with $\bm{x}$ replaced by $\bm{x}^*$.
\end{enumerate}


\subsection{Derivation of the DR weights} 
\label{drest_derive}
The DR estimator is given by $\widehat{\text{ATE}}_{\text{DR}} = \big[ \frac{1}{n}\sum_{i=1}^{n} \hat{m}_1(\bm{X}_i) + \sum_{i:Z_i=1} w^\text{base}_i \{ Y^{\text{obs}}_i - \hat{m}_1(\bm{X}_i) \} \big] - \big[ \frac{1}{n}\sum_{i=1}^{n} \hat{m}_0(\bm{X}_i) + \sum_{i:Z_i=0} w^\text{base}_i \{ Y^{\text{obs}}_i - \hat{m}_0(\bm{X}_i) \} \big]$, where $\hat{m}_1$ and $\hat{m}_0$ are obtained using MRI and the base weights are normalized. We prove that the second term of $\widehat{\text{ATE}}_{\text{DR}}$, i.e. $\frac{1}{n}\sum_{i=1}^{n} \hat{m}_0(\bm{X}_i) + \sum_{i:Z_i=0}w^{\text{base}}_i \{Y^{\text{obs}}_i - \hat{m}_0(\bm{X}_i)\}$ is of the form $\sum_{i:Z_i=0}w_i Y^{\text{obs}}_i$, where $w_i$ has the form given in Theorem 3. Now, from Proposition 2, we know that for a generic covariate profile $\bm{x}$, $\hat{m}_0(\bm{x}) = \sum_{i:Z_i=0} w^{\scriptscriptstyle \text{MRI}}_i(\bm{x}) Y^{\text{obs}}_i$, where $w^{\scriptscriptstyle \text{MRI}}_i(\bm{x}) = \frac{1}{n_c} + (\bm{x} - \bar{\bm{X}}_c)^\top \bm{S}^{-1}_c (\bm{X}_i - \bar{\bm{X}}_c)$. By linearity, we have
\begin{align}
\frac{1}{n}\sum_{i=1}^{n} \hat{m}_0(\bm{X}_i) + \sum_{i:Z_i=0}w^{\text{base}}_i (Y^{\text{obs}}_i - \hat{m}_0(\bm{X}_i)) &= \hat{\beta}_{0c} + \bm{\beta}^\top_{1c} \bar{\bm{X}} + \sum_{i:Z_i=0}w^{\text{base}}_i Y^{\text{obs}}_i - (\hat{\beta}_{0c} + \bm{\beta}^\top_{1c} \bar{\bm{X}}^{\text{base}}_{c}) \nonumber\\
&= \sum_{i:Z_i=0} \big\{ w^{\scriptscriptstyle \text{MRI}}_i(\bar{\bm{X}}) - w^{\scriptscriptstyle \text{MRI}}_i(\bar{\bm{X}}^{\text{base}}_c) + w^{\text{base}}_i \big\} Y^{\text{obs}}_i. 
\label{s1.3.1}
\end{align}
Note that $w^{\scriptscriptstyle \text{MRI}}_i(\bar{\bm{X}}) - w^{\scriptscriptstyle \text{MRI}}_i(\bar{\bm{X}}^{\text{base}}_{c}) + w^{\text{base}}_i = w^{\text{base}}_i + (\bar{\bm{X}} - \bar{\bm{X}}^{\text{base}}_{c})^\top \bm{S}^{-1}_c (\bm{X}_i - \bar{\bm{X}}_c)$. Since $w^{\text{base}}_i$s are normalized, we get
\begin{equation}
    w_i = \tilde{w}^{\text{base}}_i + (\bm{X}_i - \bar{\bm{X}}_c)^\top \bm{S}^{-1}_c (\bar{\bm{X}} - \bar{\bm{X}}^{\text{base}}_{c}).
\end{equation}

\subsection{Variance of the URI and MRI weights}
\label{var_urimri}
We first derive the variance of the URI weights in the treatment group. The variance of the weights in the control group can be derived analogously.
\begin{align}
    &\frac{1}{n_t}\sum_{i:Z_i=1}(w^{\scriptscriptstyle \text{URI}}_i - \frac{1}{n_t})^2 \nonumber\\
    &= \frac{1}{n_t}\sum_{i:Z_i=1} \big\{\frac{n}{n_c} (\bar{\bm{X}} - \bar{\bm{X}}_t)^\top (\bm{S}_t + \bm{S}_c)^{-1}(\bm{X}_i  -\bar{\bm{X}}_t) \big\}^2 \nonumber\\
    &= \frac{1}{n_t}\frac{n^2}{n^2_c}(\bar{\bm{X}} - \bar{\bm{X}}_t)^\top (\bm{S}_t + \bm{S}_c)^{-1} \big\{ \sum_{i:Z_i=1} (\bm{X}_i  -\bar{\bm{X}}_t) (\bm{X}_i  -\bar{\bm{X}}_t)^\top \big\} (\bm{S}_t + \bm{S}_c)^{-1} (\bar{\bm{X}} - \bar{\bm{X}}_t)\nonumber\\
    &= \frac{1}{n_t}\frac{n^2}{n^2_c}(\bar{\bm{X}} - \bar{\bm{X}}_t)^\top (\bm{S}_t + \bm{S}_c)^{-1} \bm{S}_t (\bm{S}_t + \bm{S}_c)^{-1} (\bar{\bm{X}} - \bar{\bm{X}}_t).
    \label{s1.4.1}
\end{align}
Similarly, the variance of the MRI weights (for the ATE case) in the treatment group can be derived as follows.
\begin{align}
    \frac{1}{n_t}\sum_{i:Z_i=1}(w^{\scriptscriptstyle \text{MRI}}_i - \frac{1}{n_t})^2 &= \frac{1}{n_t}\sum_{i:Z_i=0} \big\{ (\bar{\bm{X}} - \bar{\bm{X}}_t)^\top \bm{S}_t^{-1}  (\bm{X}_i -\bar{\bm{X}}_t)\big\}^2 \nonumber\\
    &= \frac{1}{n_t}(\bar{\bm{X}} - \bar{\bm{X}}_t)^\top \bm{S}_t^{-1} \big\{ \sum_{i:Z_i=1}(\bm{X}_i -\bar{\bm{X}}_t)(\bm{X}_i -\bar{\bm{X}}_t)^\top \big\} \bm{S}_t^{-1} (\bar{\bm{X}} - \bar{\bm{X}}_t) \nonumber\\
    &= \frac{1}{n_t}(\bar{\bm{X}} - \bar{\bm{X}}_t)^\top \bm{S}_t^{-1} (\bar{\bm{X}} - \bar{\bm{X}}_t).
    \label{s1.4.2}
\end{align}

Now, from Equations \ref{s1.4.1} and \ref{s1.4.2}, we get
\begin{align}
    &\sum_{i:Z_i=1}(w^{\scriptscriptstyle \text{URI}}_i - \frac{1}{n_t})^2 - \sum_{i:Z_i=1}(w^{\scriptscriptstyle \text{MRI}}_i - \frac{1}{n_t})^2 \nonumber\\
    &= (\bar{\bm{X}} - \bar{\bm{X}}_t)^\top \big\{\frac{n^2}{n^2_c}(\bm{S}_t + \bm{S}_c)^{-1}\bm{S}_t (\bm{S}_t + \bm{S}_c)^{-1} - \bm{S}^{-1}_t  \big\} (\bar{\bm{X}} - \bar{\bm{X}}_t)^\top \nonumber\\
    &= (\bar{\bm{X}} - \bar{\bm{X}}_t)^\top \bm{S}^{-\frac{1}{2}}_t \big\{\frac{n^2}{n^2_c}\bm{S}^{\frac{1}{2}}_t(\bm{S}_t + \bm{S}_c)^{-1}\bm{S}^{\frac{1}{2}}_t \bm{S}^{\frac{1}{2}}_t (\bm{S}_t + \bm{S}_c)^{-1} \bm{S}^{\frac{1}{2}}_t - \underline{\bm{I}}  \big\}\bm{S}^{-\frac{1}{2}}_t (\bar{\bm{X}} - \bar{\bm{X}}_t)^\top, 
    \label{s1.4.18}
\end{align}
where $\bm{S}^{\frac{1}{2}}_t$ is the symmetric square root matrix of $\bm{S}_t$, and $\bm{S}^{-\frac{1}{2}}_t := (\bm{S}^{\frac{1}{2}}_t)^{-1}$. Now, let us assume $n_t\bm{S}_t \succcurlyeq n_c \bm{S}_c$. We get,
\begin{align}
    n_t\bm{S}_t \succcurlyeq n_c \bm{S}_c &\implies n\bm{S}_t \succcurlyeq n_c(\bm{S}_t +  \bm{S}_c)\nonumber\\ 
        & \implies (\bm{S}_t + \bm{S}_c)^{-1} \succcurlyeq \frac{n_c}{n}\bm{S}^{-1}_t \nonumber\\
    & \implies  \frac{n}{n_c}\bm{S}^{\frac{1}{2}}_t(\bm{S}_t + \bm{S}_c)^{-1}\bm{S}^{\frac{1}{2}}_t \succcurlyeq \underline{\bm{I}} \nonumber\\
& \implies \big\{\frac{n}{n_c}\bm{S}^{\frac{1}{2}}_t(\bm{S}_t + \bm{S}_c)^{-1}\bm{S}^{\frac{1}{2}}_t\}^2 \succcurlyeq \underline{\bm{I}}
    \label{s1.4.19}
\end{align}
Equation \ref{s1.4.19} implies that $\frac{n^2}{n^2_c}\bm{S}^{\frac{1}{2}}_t(\bm{S}_t + \bm{S}_c)^{-1}\bm{S}^{\frac{1}{2}}_t \bm{S}^{\frac{1}{2}}_t (\bm{S}_t + \bm{S}_c)^{-1} \bm{S}^{\frac{1}{2}}_t - \underline{\bm{I}}$ is non-negative definite, and hence from Equation \ref{s1.4.18} we get, $\sum_{i:Z_i=1}(w^{\scriptscriptstyle \text{URI}}_i - \frac{1}{n_t})^2 \geq \sum_{i:Z_i=1}(w^{\scriptscriptstyle \text{MRI}}_i - \frac{1}{n_t})^2$. Thus, when $n_t\bm{S}_t \succcurlyeq n_c \bm{S}_c$, the variance of the URI weights in the treatment group is no less than that of the MRI weights. By similar calculations it follows that in this case, the variance of the URI weights in the treatment group is no greater than that of the MRI weights. The inequalities are reversed when $n_t\bm{S}_t \preccurlyeq n_c \bm{S}_c$.

We now compare the total variance of the URI and MRI weights across all $n$ units in the sample. By similar calculations, we get
\begin{align}
    \sum_{i=1}^{n}(w^{\scriptscriptstyle \text{MRI}}_i - \frac{2}{n})^2 - \sum_{i=1}^{n}(w^{\scriptscriptstyle \text{URI}}_i - \frac{2}{n})^2 &= (\bar{\bm{X}}_t - \bar{\bm{X}}_c)^\top \big\{\frac{1}{n^2}(n^2_c \bm{S}^{-1}_t + n^2_t \bm{S}^{-1}_c) - (\bm{S}_t + \bm{S}_c)^{-1}  \big\} (\bar{\bm{X}}_t - \bar{\bm{X}}_c)
\end{align}
We will now show that $\frac{1}{n^2}(n^2_c \bm{S}^{-1}_t + n^2_t \bm{S}^{-1}_c) \succcurlyeq (\bm{S}_t + \bm{S}_c)^{-1}$, which is equivalent to showing $\frac{1}{n^2}(n^2_c \bm{S}^{-1}_t + n^2_t \bm{S}^{-1}_c) (\bm{S}_t + \bm{S}_c) \succcurlyeq  \underline{\bm{I}}$. 
We now use the following lemma.
\begin{lemma}
For a $k \times k$ non-negative definite matrix $\bm{A}$, $\frac{1}{2}(\bm{A}+\bm{A}^{-1}) \succcurlyeq \underline{\bm{I}}$.
\label{lemma_1.1}
\end{lemma}
\noindent \textit{Proof of Lemma \ref{lemma_1.1}.} Using the spectral decomposition of $\bm{A}$, we can write $\bm{A} = \bm{P}\bm{\Lambda}\bm{P}^\top$, where $\bm{P}$ is an orthogonal matrix and $\Lambda$ is the diagonal matrix of the (ordered) eigenvalues of $\bm{A}$. Therefore, $\frac{1}{2}(\bm{A} + \bm{A}^{-1})  =  \bm{P}\frac{1}{2}(\bm{\Lambda} + \bm{\Lambda}^{-1})\bm{P}^\top$. Therefore, if $(\lambda_1,...,\lambda_k)$ are the eigenvalues of $A$, the eigenvalues of $\frac{1}{2}(\bm{A} + \bm{A}^{-1})$ are $(\lambda_1 + \frac{1}{\lambda_1},...,\lambda_k+ \frac{1}{\lambda_k})$. By AM-GM inequality, $\forall i \in \{1,...,k\}$, $\frac{1}{2}(\lambda_i + \frac{1}{\lambda_i}) \geq 1$. So the eigenvalues of $\frac{1}{2}(\bm{A} + \bm{A}^{-1}) - \underline{\bm{I}}$ are non-negative, which completes the proof.

Now, let $A = n_t^{-1}n_c \bm{S}_t^{-\frac{1}{2}}\bm{S}_c\bm{S}_t^{-\frac{1}{2}}$, which is non-negative definite. Lemma \ref{lemma_1.1} implies $\frac{1}{2}(n_t^{-1}n_c \bm{S}_t^{-\frac{1}{2}}\bm{S}_c\bm{S}_t^{-\frac{1}{2}} + n_c^{-1}n_t \bm{S}_t^{\frac{1}{2}}\bm{S}^{-1}_c\bm{S}_t^{\frac{1}{2}}) \succcurlyeq \underline{\bm{I}}$. Finally, we note that
\begin{align}
    \frac{1}{2}(n_t^{-1}n_c \bm{S}_t^{-\frac{1}{2}}\bm{S}_c\bm{S}_t^{-\frac{1}{2}} + n_c^{-1}n_t \bm{S}_t^{\frac{1}{2}}\bm{S}^{-1}_c\bm{S}_t^{\frac{1}{2}}) \succcurlyeq \underline{\bm{I}} &\implies \frac{1}{2}(n_t^{-1}n_c \bm{S}_c\bm{S}_t^{-1} + n_c^{-1}n_t \bm{S}_t\bm{S}_c^{-1}) \succcurlyeq \underline{\bm{I}} \nonumber\\
    & \implies n^2_c \bm{S}_c\bm{S}^{-1}_t + n^2_t \bm{S}_t\bm{S}^{-1}_c \succcurlyeq 2n_tn_c \nonumber\\
    & \implies \frac{1}{n^2}(n^2_c \bm{S}^{-1}_t + n^2_t \bm{S}^{-1}_c) (\bm{S}_t + \bm{S}_c) \succcurlyeq  \underline{\bm{I}}
\end{align}
This proves that the variance of the MRI weights in the full-sample is no less than that of the URI weights.
\subsection{Bias of the URI and MRI estimators}
Consider a generic H\'{a}jek estimator $T = \sum_{i:Z_i=1}w_i Y^{\text{obs}}_i - \sum_{i:Z_i=0}w_i Y^{\text{obs}}_i$ of the ATE, where the weights are normalized within each treatment group. Under unconfoundedness, the bias of $T$ due to imbalances on the observed covariates is completely removed if the weights satisfy $\sum_{i:Z_i=1}w_i m_1(\bm{X}_i) = n^{-1}\sum_{i=1}^{n} m_1(\bm{X}_i)$ and $\sum_{i:Z_i=0}w_i m_0(\bm{X}_i) = n^{-1}\sum_{i=1}^{n} m_0(\bm{X}_i)$. 
As a special case, when both $m_1(\cdot)$ and $m_0(\cdot)$ are linear in $\bm{X}_i$, balancing the mean of $\bm{X}_i$ relative to $\bar{\bm{X}}$ suffices to remove the bias of $T$. 
Now, if $\sum_{i:Z_i=1}w_i\bm{X}_i = \sum_{i:Z_i=0}w_i\bm{X}_i = \bm{X}^*$, then ${E}(T\mid \bm{Z},\underline{\bm{X}}) = \text{CATE}(\bm{X}^*)$. This implies that the URI estimates the average treatment effect on a population characterized by the profile $\bm{X}^{*{\scriptscriptstyle \text{URI}}}$, whereas MRI estimates the average treatment effect on a population characterized by the profile $\bm{X}^{*{\scriptscriptstyle \text{MRI}}}$. Now, treatment effect homogeneity implies $\text{CATE}(\bm{x}) = \text{ATE}$ for all $\bm{x} \in \text{supp}(\bm{X}_i)$. 
Thus, under linearity and treatment effect homogeneity, the disparity between the implied target population under URI and the intended target population does not matter and the URI estimator is unbiased for the ATE.  
However, if CATE($\bm{x}$) is a non-trivial function of $\bm{x}$, then ${E}\{\text{CATE}(\bm{X}^{*{\scriptscriptstyle \text{URI}}})\} \neq \text{ATE}$ in general and the URI estimator is biased for the ATE, despite balancing the mean of $\bm{X}_i$ exactly. 
On the other hand, as long as $m_0(\cdot)$ and $m_1(\cdot)$ are linear in $\bm{X}_i$, ${E}\{\text{CATE}(\bar{\bm{X}})\} = \text{ATE}$ and the MRI estimator is unbiased for the ATE. 
However, if $m_0(\cdot)$ and $m_1(\cdot)$ are linear on some other transformations of $\bm{X}_i$, both URI and MRI weights can produce biased estimators, since the implied weights are not guaranteed to yield exact mean balance on these transformations.

\subsection{URI and MRI weights under no-intercept model}
Let $\bm{y}$, $\bm{y}_t$, and $\bm{y}_c$ be the vector of observed outcomes in the full-sample, treatment group, and control group, respectively. In the URI approach with a no-intercept model, we fit the regression model $Y^{\text{obs}}_i = \bm{\beta}^\top \bm{X}_i + \tau Z_i +  \epsilon_i$. Let $\bm{\mathcal{P}}$ be the projection matrix onto the column space of $\underline{\bm{X}}$. By the Frisch–Waugh–Lovell theorem (\citealt{frisch1933partial},\citealt{lovell1963seasonal}), the OLS estimator of $\tau$ can be written as,
\begin{equation}
    \hat{\tau}^{\text{OLS}} = \frac{\bm{Z}^\top (\underline{\bm{I}} - \bm{\mathcal{P}})\bm{y}}{\bm{Z}^\top (\underline{\bm{I}} - \bm{\mathcal{P}})\bm{Z}} = \bm{l}^\top \bm{y}.
    \label{s1.5.1}
\end{equation}
where $\bm{l} = (l_1,...,l_n)^\top = \frac{(\underline{\bm{I}} - \bm{\mathcal{P}})\bm{Z}}{\bm{Z}^\top (\underline{\bm{I}} - \bm{\mathcal{P}})\bm{Z}}$. So, $\hat{\tau}^{\text{OLS}}$ can be written as $\hat{\tau} = \sum_{i:Z_i=1}w^{\scriptscriptstyle \text{URI}}_i Y^{\text{obs}}_i - \sum_{i:Z_i=0}w^{\scriptscriptstyle \text{URI}}_i Y^{\text{obs}}_i$, where $w^{\scriptscriptstyle \text{URI}}_i = (2Z_i-1)l_i$. We observe that $\sum_{i:Z_i=1}w^{\scriptscriptstyle \text{URI}}_i = \bm{l}^\top \bm{Z} = 1$. So in this case, the URI weights are normalized in the treatment group. However, $\sum_{i:Z_i=0}w^{\scriptscriptstyle \text{URI}}_i$ may not be equal to 1 in general. Also, we note that $\sum_{i:Z_i=1}w^{\scriptscriptstyle \text{URI}}_i\bm{X}_i - \sum_{i:Z_i=0}w^{\scriptscriptstyle \text{URI}}_i\bm{X}_i = \underline{\bm{X}}^\top \bm{l} = \bm{0}$, since $\underline{\bm{X}}^\top \bm{\mathcal{P}} = \underline{\bm{X}}^\top$. So, the weighted \textit{sums} of the covariates are the same in the treatment and the control group. However, the weighted \textit{means} of the covariates are not guaranteed to be the same.

Similarly, in the MRI approach, we fit the model $Y^{\text{obs}}_i = \bm{\beta}_t\bm{X}_i + \epsilon_{it}$ in the treatment group, and $Y^{\text{obs}}_i = \bm{\beta}^\top_c\bm{X}_i + \epsilon_{ic}$ in the control group. 
For a fixed profile $\bm{x} \in \mathbb{R}^k$, $\hat{m}_1(\bm{x}) = \hat{\bm{\beta}}^\top_{t}\bm{x} = \bm{w}^\top_t \bm{y}_t$, where $\bm{w}_t = \underline{\bm{X}}_t(\underline{\bm{X}}^\top_t \underline{\bm{X}}_t)^{-1} \bm{x}$. Thus, $\hat{m}_1(\bm{x})$ can be written as $\hat{m}_1(\bm{x}) = \sum_{i:Z_i=1}w^{\scriptscriptstyle \text{MRI}}_i(\bm{x}) Y^{\text{obs}}_i$. Note that here the weights do not necessarily sum to one. However, $\sum_{i:Z_i=1}w^{\scriptscriptstyle \text{MRI}}_i(\bm{x}) \bm{X}_i = \underline{\bm{X}}^\top_t \bm{w}_t = \bm{x}$. Therefore, the weighted \textit{sum} of the covariates in the treatment group is balanced relative to the target profile, but the corresponding weighted \textit{mean} of the covariates is imbalanced in general.


\subsection{Asymptotic properties of URI}

\begin{theorem}{\label{uri_conv}}
Let $w^{{\scriptscriptstyle \text{URI}}}_{\bm{x}}$ be the URI weight of a unit with covariate vector $\bm{x}$. Assume that $p^2 \Var(\bm{X}_i\mid Z_i=1) = (1-p)^2 \Var(\bm{X}_i\mid Z_i=0)$, where $p = P(Z_i=1)$. Then
\begin{enumerate}[label = (\alph*)]

\item For each treated unit, $n w^{{\scriptscriptstyle \text{URI}}}_{\bm{x}} \xrightarrow[n \to \infty]{P} \frac{1}{e(\bm{x})}$ for all $\bm{x} \in \text{supp}(\bm{X}_i)$ if and only if the propensity score is an inverse-linear function of the covariates; i.e., $e(\bm{x}) = \frac{1}{\alpha_0 + \bm{\alpha}^\top_1 \bm{x}}$, $\alpha_0 \in \mathbb{R}$, $\bm{\alpha}_1 \in \mathbb{R}^k$. Moreover, if $\sup\limits_{\bm{x} \in \text{supp}(\bm{X}_i)}\norm{\bm{x}}_2 < \infty$, $\sup \limits_{\bm{x} \in \text{supp}(\bm{X}_i)}\mid nw^{{\scriptscriptstyle \text{URI}}}_{\bm{x}} - \frac{1}{e(\bm{x})}\mid  \xrightarrow[n \to \infty]{P} 0$.

\item Similarly, for each control unit, $n w^{{\scriptscriptstyle \text{URI}}}_{\bm{x}} \xrightarrow[n \to \infty]{P} \frac{1}{1-e(\bm{x})}$ if and only if $1-e(\bm{x})$ is inverse linear function of the covariates, and the convergence is uniform if $\sup\limits_{\bm{x} \in \text{supp}(\bm{X}_i)}\norm{\bm{x}}_2 < \infty$.

\end{enumerate}

\end{theorem}
\textit{Proof of Theorem \ref{uri_conv}.} 
Let $\bm{\mu}_t = {E}(\bm{X}_i\mid Z_i=1)$, $\bm{\mu}_c = {E}(\bm{X}_i\mid Z_i=0)$, $\bm{\mu} = {E}(\bm{X}_i)$, and $\bm{\Sigma}_t = \Var(\bm{X}_i\mid Z_i=1)$, $\bm{\Sigma}_c = \Var(\bm{X}_i\mid Z_i=0)$. By WLLN and Slutsky's theorem, we have $\bar{\bm{X}}_t = \frac{\frac{1}{n}\sum_{i=1}^{n}Z_i\bm{X}_i}{\frac{1}{n}\sum_{i=1}^{n} Z_i} \xrightarrow[n \to \infty]{P} \frac{{E}(Z_i \bm{X}_i)}{p} = \bm{\mu}_t$. Similarly, we have $\frac{\bm{S}_t}{n_t} \xrightarrow[n \to \infty]{P} \bm{\Sigma}_t$ and $\frac{\bm{S}_c}{n_c} \xrightarrow[n \to \infty]{P} \bm{\Sigma}_c$. Now, we consider a treated unit with covariate vector $\bm{x} \in \text{supp}(\bm{X}_i)$. By continuous mapping theorem, we have 
\begin{align}
nw^{\scriptscriptstyle \text{URI}}_{\bm{x}} &= \frac{n}{n_t} + \frac{n}{n_c}(\bm{x} - \bar{\bm{X}}_t)^\top \big(\frac{\bm{S}_t}{n_t}\frac{n_t}{n} + \frac{\bm{S}_c}{n_c}\frac{n_c}{n}\big)^{-1} (\bar{\bm{X}} - \bar{\bm{X}}_t) \nonumber\\
&\xrightarrow[n \to \infty]{P} \frac{1}{p} \Big[ 1 + \frac{p}{1-p}(\bm{x} - \bm{\mu}_t)^\top \{p\bm{\Sigma}_t + (1-p)\bm{\Sigma}_c\}^{-1} (\bm{\mu} - \bm{\mu}_t)  \Big].
\label{s1.6.1}
\end{align}
Under the assumption that $p^2 \bm{\Sigma}_t = (1-p)^2 \bm{\Sigma}_c$, the RHS of Equation \ref{s1.6.1} boils down to $\frac{1}{p} \big\{ 1 + (\bm{x} - \bm{\mu}_t)^\top \bm{\Sigma}^{-1}_t (\bm{\mu} - \bm{\mu}_t)  \big\}$, which is same as the probability limit of $nw^{\scriptscriptstyle \text{MRI}}_{\bm{x}}$ (see the proof of Theorem 1). Thus, when $p^2 \bm{\Sigma}_t = (1-p)^2 \bm{\Sigma}_c$, the URI and MRI weights are asymptotically equivalent. The rest of the proof follows from the proof of Theorem 1. 

\begin{lemma}
Let the true propensity score be linear on the covariates, i.e., $e(\bm{x}) = a_0 + \bm{a}^\top_1\bm{x}$ for some constants $a_0 \in \mathbb{R}$, $\bm{a}_1 \in \mathbb{R}^k$. Then $$ \bm{a}_1 = \frac{p(1-p)}{1+ p(1-p)c} \bm{A}^{-1}(\bm{\mu}_t - \bm{\mu}_c),\hspace{0.1cm} a_0 = p - \bm{a}_1^\top \bm{\mu},$$
where $\bm{A} = p\bm{\Sigma}_t + (1-p)\bm{\Sigma}_c$, and $c = (\bm{\mu}_t - \bm{\mu}_c)^\top \bm{A}^{-1} (\bm{\mu}_t - \bm{\mu}_c)$. Here $p,\bm{\mu}, \bm{\mu}_t, \bm{\Sigma}_t, \bm{\Sigma}_c$ are the same as in the proof of Theorem \ref{uri_conv}.  
\label{linear_propensity}
\end{lemma}
\textit{Proof of Lemma \ref{linear_propensity}.} Since ${E}(e(\bm{X}_i)) = p$, we have $a_0 = p - \bm{a}^\top \bm{\mu}$. Next, expanding the identity ${E}((\bm{X}_i - \bm{\mu}_t)e(\bm{X}_i)) = 0$, it is straightforward to show that 
\begin{equation}
    \bm{a}_1 = p(1-p) \bm{\Sigma}^{-1}(\bm{\mu}_t - \bm{\mu}_c),
    \label{s1.6.2}
\end{equation}
where $\bm{\Sigma} = \Var(\bm{X}_i)$. Moreover, by conditioning on $Z_i$, we can decompose $\Var(\bm{X}_i)$ as 
\begin{equation}
\bm{\Sigma} = (p\bm{\Sigma}_t + (1-p)\bm{\Sigma}_c) + p(1-p)(\bm{\mu}_t - \bm{\mu}_c)(\bm{\mu}_t - \bm{\mu}_c)^\top = \bm{A} + p(1-p)(\bm{\mu}_t - \bm{\mu}_c)(\bm{\mu}_t - \bm{\mu}_c)^\top.    
\label{s1.6.3}
\end{equation}
Applying the Sherman-Morrison-Woodbury formula (\citealt{sherman1950adjustment}, \citealt{woodbury1950inverting}), we can write the inverse of $\bm{\Sigma}$ as 
\begin{equation}
    \bm{\Sigma}^{-1} = \bm{A}^{-1} - \frac{p(1-p) \bm{A}^{-1}(\bm{\mu}_t - \bm{\mu}_c)(\bm{\mu}_t - \bm{\mu}_c)^\top \bm{A}^{-1}}{1+p(1-p)c}.
    \label{s1.6.4}
\end{equation}
Substituting the expression of $\bm{\Sigma}^{-1}$ in \ref{s1.6.2}, it follows that 
\begin{equation}
    \bm{a}_1 = \frac{p(1-p)}{1+ p(1-p)c} \bm{A}^{-1}(\bm{\mu}_t - \bm{\mu}_c).
\end{equation}
This completes the proof of the Lemma.

\begin{theorem}{\label{dr_uri}}
The URI estimator for the ATE is consistent if any of the following conditions holds.
\begin{enumerate}[label=(\roman*)]
\item $m_0(\bm{x})$ is linear, $e(\bm{x})$ is inverse linear, and $p^2 \Var(\bm{X}_i\mid Z_i=1) = (1-p)^2 \Var(\bm{X}_i\mid Z_i=0)$.
\item $m_1(\bm{x})$ is linear, $1-e(\bm{x})$ is inverse linear, and $p^2 \Var(\bm{X}_i\mid Z_i=1) = (1-p)^2 \Var(\bm{X}_i\mid Z_i=0)$.
\item Both $m_1(\bm{x})$ and $m_0(\bm{x})$ are linear and $p^2 \Var(\bm{X}_i\mid Z_i=1) = (1-p)^2 \Var(\bm{X}_i\mid Z_i=0)$.
\item $e(\bm{x})$ is a constant function of $\bm{x}$.
\item Both $m_1(\bm{x})$ and $m_0(\bm{x})$ are linear and $m_1(\bm{x}) - m_0(\bm{x})$ is a constant function.
\item $m_1(\bm{x})- m_0(\bm{x})$ is a constant function and $e(\bm{x})$ is linear in $\bm{x}$.
\end{enumerate}
\end{theorem}

\noindent \textit{Proof of Theorem \ref{dr_uri}.} 
 Let $p,\bm{\mu}, \bm{\mu}_t, \bm{\Sigma}_t, \bm{\Sigma}_c$ be defined as in the proof of Theorem \ref{uri_conv}. By similar calculations as in the proof of Theorem \ref{uri_conv} we have, 
\begin{align}
\sum_{i:Z_i=1}w^{\scriptscriptstyle \text{URI}}_i Y^{\text{obs}}_i &= \bar{Y}_t + \frac{n}{n_c} (\bar{\bm{X}} - \bar{\bm{X}}_t)^\top \big(\frac{\bm{S}_t + \bm{S}_c}{n} \big)^{-1} \big\{ \frac{1}{n} \sum_{i:Z_i=1} (\bm{X}_i - \bar{\bm{X}}_t)Y^{\text{obs}}_i  \big\} \nonumber \\
&\xrightarrow[n \to \infty]{P} {E} \Big(m_1(\bm{X}_i)e(\bm{X}_i) \big\{\frac{1}{p} +  \frac{1}{1-p} (\bm{\mu} - \bm{\mu}_t)^\top (p\bm{\Sigma}_t + (1-p)\bm{\Sigma}_c)^{-1}(\bm{X}_i - \bm{\mu}_t)\big\} \Big) .
\label{1.6.2}
\end{align}

First, if $p^2 \bm{\Sigma}_t = (1-p)^2 \bm{\Sigma}_c$, the right hand side of Equation 1 becomes ${E}\Big(\frac{m_1(\bm{X}_i)e(\bm{X}_i)}{p} \big\{1+ (\bm{\mu} - \bm{\mu}_t)^\top \bm{\Sigma}^{-1}_t(\bm{x} - \bm{\mu}_t) \big\}    \Big)$, which is same as the probability limit of $\sum_{i:Z_i=1}w^{\scriptscriptstyle \text{MRI}}_i Y^{\text{obs}}_i$. Similarly, when $p^2 \bm{\Sigma}_t = (1-p)^2 \bm{\Sigma}_c$, we can show that $\sum_{i:Z_i=0}w^{\scriptscriptstyle \text{URI}}_i Y^{\text{obs}}_i$ has the same probability limit as $\sum_{i:Z_i=0} w^{\scriptscriptstyle \text{MRI}}_i Y^{\text{obs}}_i$. This observation, along with conditions Theorem 2, proves consistency of the URI estimator under parts (i), (ii) and (iii) of Theorem \ref{dr_uri}. Second, if the propensity score is constant, $\bm{\mu} = \bm{\mu}_t$ and the right hand side of Equation \ref{1.6.2} becomes ${E}(m_1(\bm{X}_i))$, which equals ${E}(Y_i(1))$. Similarly, in this case, the probability limit of $\sum_{i:Z_i=0} w^{\scriptscriptstyle \text{URI}}Y^{\text{obs}}_i$ becomes ${E}(Y_i(0))$. This proves consistency of the URI estimator under \textit{(iv)}. Third, by consistency of OLS estimators of regression coefficients under well-specified model, the URI estimator is consistent for the ATE under part (v).

Finally, let $e(\bm{x})= a_0 + \bm{a}_1^\top \bm{x}$. Using the notation in Lemma \ref{linear_propensity}, we know that $ \bm{a}_1 = \frac{p(1-p)}{1+ p(1-p)c} \bm{A}^{-1}(\bm{\mu}_t - \bm{\mu}_c)$, $a_0 = p - \bm{a}_1^\top \bm{\mu}$. Let $d = {E}(e(\bm{X}_i)(1-e(\bm{X}_i))$. Using the expressions of $a_0$ and $\bm{a}_1$, it is straightforward to show that
\begin{equation}
d = \frac{p(1-p)}{1+p(1-p)c}.
    \label{1.6.3}
\end{equation}
This implies, 
\begin{equation}
    1-e(\bm{x}) = d \big\{ \frac{1}{p} + \frac{1}{1-p}(\bm{\mu} - \bm{\mu}_t) \bm{A}^{-1}(\bm{x} - \bm{\mu}_t) \big\}
    \label{1.6.4}
\end{equation}
Equations \ref{1.6.2} and \ref{1.6.4} imply,
\begin{equation}
    \sum_{i:Z_i=1}w^{\scriptscriptstyle \text{URI}}_i Y^{\text{obs}}_i \xrightarrow[n \to \infty]{P} \frac{{E}(e(\bm{X}_i)\{1-e(\bm{X}_i)\}m_1(\bm{X}_i))}{{E}(e(\bm{X}_i)\{1-e(\bm{X}_i)\}]}.
    \label{1.6.5}
\end{equation}
By similar calculations for the control group, we get
\begin{equation}
    \sum_{i:Z_i=0}w^{\scriptscriptstyle \text{URI}}_i Y^{\text{obs}}_i \xrightarrow[n \to \infty]{P} \frac{{E}(e(\bm{X}_i)\{1-e(\bm{X}_i)\}m_0(\bm{X}_i))}{{E}(e(\bm{X}_i)\{1-e(\bm{X}_i)\}]}.
    \label{1.6.6}
\end{equation}
Equations \ref{1.6.5} and \ref{1.6.6} imply that, when the propensity score is linear on the covariates,
\begin{equation}
    \sum_{i:Z_i=1}w^{\scriptscriptstyle \text{URI}}_i Y^{\text{obs}}_i -  \sum_{i:Z_i=0}w^{\scriptscriptstyle \text{URI}}_i Y^{\text{obs}}_i \xrightarrow[n \to \infty]{P} \frac{{E}(e(\bm{X}_i)\{1-e(\bm{X}_i)\}\{m_1(\bm{X}_i) - m_0(\bm{X}_i)\} )}{{E}(e(\bm{X}_i)\{1-e(\bm{X}_i)\})}.
    \label{1.6.7}
\end{equation}
Note that this limiting representation of the URI estimator is equivalent to that in \cite{aronow2016does}. Now, the consistency of the URI estimator under condition (vi) follows from Equation \ref{1.6.7} by noting that if $m_1(\bm{x}) - m_0(\bm{x}) = \tau$ for all $\bm{x} \in \text{supp}(\bm{X}_i)$, the RHS of Equation \ref{1.6.7} equals $\tau$.

\section{Proofs of propositions and theorems}

\subsection{Proof of Theorem 3}

For $\bm{\delta} = 0$, the Lagrangian of the optimization problem is given by 
\begin{equation}
L(\bm{w}, \lambda_1, \bm{\lambda}_2) = \sum_{i: Z_i=0}\frac{(w_i - \tilde{w}^\text{base}_i)^2}{w^\text{scale}_i} + \lambda_1 (\sum_{i:Z_i=0}w_i - 1) + \bm{\lambda}_2^\top (\sum_{i:Z_i=0} w_i \bm{X}_i - \bm{X}^*).
\label{6.1_1}
\end{equation}

Computing the partial derivatives $\frac{\partial L}{\partial \bm{w}}$, $\frac{\partial L}{\partial \lambda_1}$, $\frac{\partial L}{\partial \bm{\lambda}_2}$ and equating them to zero, we get the following equations:

\begin{equation} 
w_i = \tilde{w}^\text{base}_i - w^\text{scale}_i \frac{\lambda_1 + \bm{\lambda}_2^\top \bm{X}_i}{2} \hspace{0.2cm} \text{for all $i:Z_i=0$.}
\label{6.1_2}
\end{equation}
\begin{equation}
\sum_{i: Z_i=0} w_i=1
\label{6.1_3}
\end{equation}
\begin{equation}
\sum_{i: Z_i=0} w_i \bm{X}_i^\top = \bm{X}^{*\top}
\label{6.1_4}
\end{equation}
Substituting the expression of $w_i$ from Equation \ref{6.1_2} in Equation \ref{6.1_3}, we get,
\begin{equation} 
\lambda_1  + \bm{\lambda}^\top_2 \bar{\bm{X}}^{\text{scale}}_c = 0\\
\iff \lambda_1 =  - \bm{\lambda}^\top_2 \bar{\bm{X}}^{\text{scale}}_c
\end{equation}

Substituting the expression of $w_i$ from Equation \ref{6.1_2} in Equation \ref{6.1_4}, we get,
\begin{align}
&\bar{\bm{X}}^{\text{base}\top}_c - \frac{1}{2} \Big[(\sum_{i:Z_i=0}w^\text{scale}_i)\lambda_1 \bar{\bm{X}}^{\text{scale} \top}_c + 
\bm{\lambda}^\top_2 \{\frac{\bm{S}^{\text{scale}}_c}{n_c} + \bar{\bm{X}}^{\text{scale}}_c\bar{\bm{X}}^{\text{scale} \top}_c (\sum_{i:Z_i=0}w^\text{scale}_i) \} \Big] = \bm{X}^{*\top} \nonumber\\
&\iff  \bar{\bm{X}}^{\text{base}\top}_c - \frac{1}{2} \bm{\lambda}_2^\top \frac{\bm{S}^{\text{scale}}_c}{n_c} = \bm{X}^{*\top} \nonumber\\
&\iff \bm{\lambda}_2 = 2 \Big(\frac{\bm{S}^{\text{scale}}_c}{n_c} \Big)^{-1} (\bar{\bm{X}}^{\text{base}}_c - \bm{X}^*)
\end{align}
Substituting $\lambda_1$ and $\bm{\lambda}_2$ in Equation \ref{6.1_2}, we get the resulting expression of $w_i$.

\noindent The corresponding results for the WURI, WMRI, and DR weights follow from the derivations in Sections \ref{wuri_derive}, \ref{wmri_derive}, and \ref{drest_derive} of the Supplementary Materials, respectively.



\subsection{Proof of Proposition 1}

The proof follows from setting $w^{\text{base}}_i = \frac{1}{n}$ in the derivation of WURI weights in Section \ref{wuri_derive} of the Supplementary Materials.

\subsection{Proof of Proposition 2} 
The proof follows from setting $w^{\text{base}}_i = \frac{1}{n_t}$ for all $i:Z_i=1$, $w^{\text{base}}_i = \frac{1}{n_c}$ for all $i:Z_i=0$ in the derivation of WMRI weights in Section \ref{wmri_derive} of the Supplementary Materials.

\subsection{Proof of Proposition 3}
Parts (a), (b), and (e) of Proposition 3 directly follows from Theorem 3. Part (d) is a direct consequence of the closed form expression of the weights, given in Propositions 3.1 and 3.2 (see also Section 5.2 of the paper). Part (b) follows from Section \ref{var_urimri} of the Supplementary Materials.
\subsection{Proof of Theorem 1} 

Let $p = P(Z_i=1)$, $\bm{\mu}_t = {E}(\bm{X}_i\mid Z_i=1)$, $\bm{\mu} = {E}(\bm{X}_i)$, and $\bm{\Sigma}_t = \Var(\bm{X}_i\mid Z_i=1)$. We first show that when $e(\bm{x})$ is of the form $\frac{1}{\alpha_0 + \bm{\alpha}^\top_1 \bm{x}}$, then
\begin{equation}
\alpha_0 + \bm{\alpha}^\top_1 \bm{x} = \frac{1}{p} \big\{1 + (\bm{x} - \bm{\mu}_t)^\top \bm{\Sigma}^{-1}_t (\bm{\mu} - \bm{\mu}_t)\big\}
\label{4.2.1}
\end{equation}
Denoting $\bm{b}_1 = p \bm{\Sigma}_t \bm{\alpha}_1$ and $b_0 = \alpha_0 + \frac{1}{p}\bm{b}^\top_1 \bm{\Sigma}^{-1}_t \bm{\mu}_t$, we have
\begin{equation}
\frac{1}{e(\bm{x})} = b_0  + \frac{1}{p}\bm{b}^\top_1 \bm{\Sigma}^{-1}_t (\bm{x} - \bm{\mu}_t)
\label{4.2.2}
\end{equation}
It is enough to show $b_0 = \frac{1}{p}$ and $\bm{b}_1 = \bm{\mu} - \bm{\mu}_t$. Now,
\begin{align}
\bm{\mu}_t &= \frac{{E}(\bm{X}_i e(\bm{X}_i))}{p}\label{4.2.3}\\
\implies \frac{1}{p} \bm{b}^\top_1 \bm{\Sigma}^{-1} \bm{\mu}_t &= \frac{1}{p} + \frac{1}{p} {E} \big[ \frac{\frac{1}{p}\bm{b}^\top_1 \bm{\Sigma}^{-1}_t \bm{\mu}_t - b_0}{b_0 + \frac{1}{p}\bm{b}^\top_1 \bm{\Sigma}^{-1}_t (\bm{X}_i - \bm{\mu}_t )} \big] \label{4.2.4}\\
\implies \frac{1}{p} \bm{b}^\top_1 \bm{\Sigma}^{-1} \bm{\mu}_t &= \frac{1}{p} + \frac{1}{p} \big(\frac{1}{p}\bm{b}^\top_1 \bm{\Sigma}^{-1}_t \bm{\mu}_t - b_0  \big) p \label{4.2.5}\\
\implies b_0 &= \frac{1}{p}.
\label{4.2.6}
\end{align}
Here Equation \ref{4.2.3} holds by definition of conditional expectation and law of iterated expectations. Equation \ref{4.2.4} is obtained by multiplying both sides of Equation \ref{4.2.3} by $\frac{1}{p}\bm{b}^\top_1 \bm{\Sigma}^{-1}_t$ and applying Equation \ref{4.2.2}. Equation \ref{4.2.5} holds since $p = {E}(e(\bm{X}_i))$. 
Similarly,
\begin{align}
\bm{\Sigma}_t &= \frac{{E}((\bm{X}_i - \bm{\mu}_t)(\bm{X}_i - \bm{\mu}_t)^\top e(\bm{X}_i))}{p} \label{4.2.7}\\
\implies \frac{\bm{b}^\top_1}{p} &= \frac{1}{p}{E}((\bm{X}_i - \bm{\mu}_t)^\top) - \frac{1}{p^2}{E}((\bm{X}_i - \bm{\mu}_t)^\top e(\bm{X}_i)) \label{4.2.8}\\
\implies \bm{b}_1 &= \bm{\mu} - \bm{\mu}_t. 
\label{4.2.9}
\end{align}
Here Equation \ref{4.2.7} holds by definition of conditional expectation and law of iterated expectation. Equation \ref{4.2.8} is obtained by multiplying both sides of Equation \ref{4.2.7} by $\frac{\bm{b}^\top_1 \bm{\Sigma}^{-1}_t}{p}$ and applying Equation \ref{4.2.2}. Equation \ref{4.2.9} holds since ${E}((\bm{X}_i - \bm{\mu}_t)^\top e(\bm{X}_i)) = 0$. This proves Equation \ref{4.2.1}. 

Now, by WLLN and Slutsky's theorem, we have $\bar{\bm{X}}_t = \frac{\frac{1}{n}\sum_{i=1}^{n}Z_i\bm{X}_i}{\frac{1}{n}\sum_{i=1}^{n} Z_i} \xrightarrow[n \to \infty]{P} \frac{{E}(Z_i \bm{X}_i)}{p} = \bm{\mu}_t$. Similarly, we have $\frac{\bm{S}_t}{n_t} \xrightarrow[n \to \infty]{P} \bm{\Sigma}_t$. Now, we consider a treated unit with covariate vector $\bm{x} \in \text{supp}(\bm{X}_i)$. By continuous mapping theorem, we have 
\begin{equation}
nw^{\scriptscriptstyle \text{MRI}}_{\bm{x}} = \frac{n}{n_t} + \frac{n}{n_t}(\bm{x} - \bar{\bm{X}}_t)^\top \big(\frac{\bm{S}_t}{n_t}\big)^{-1} (\bar{\bm{X}} - \bar{\bm{X}}_t) \xrightarrow[n \to \infty]{P} \frac{1}{p} \big\{ 1 + (\bm{x} - \bm{\mu}_t)^\top \bm{\Sigma}^{-1}_t (\bm{\mu} - \bm{\mu}_t)  \big\}.
\label{4.2.10}
\end{equation}
This proves pointwise convergence of the MRI weights for a treated unit. To prove uniform convergence, we assume $\sup\limits_{\bm{x} \in Supp(\bm{X}_i)}\mid \mid \bm{x}\mid \mid _2 <\infty$. 
\begin{align}
\sup\limits_{\bm{x} \in \text{supp}(\bm{X}_i)}\mid nw^{\scriptscriptstyle \text{MRI}}_{\bm{x}} - \frac{1}{e(\bm{x})}\mid  &\leq \mid \frac{n}{n_t} - \frac{1}{p}\mid  + \mid \frac{n}{n_t}\bar{\bm{X}}^\top_t \big(\frac{\bm{S}_t}{n_t} \big)^{-1} (\bar{\bm{X}} - \bar{\bm{X}}_t) - \frac{1}{p} \bm{\mu}^\top_t\bm{\Sigma}^{-1}_t(\bm{\mu} - \bm{\mu}_t)\mid  \nonumber\\
& + \sup\limits_{\bm{x} \in \text{supp}(\bm{X}_i)}\mid \big\{\frac{n}{n_t}(\bar{\bm{X}} - \bar{\bm{X}}_t)^\top \big(\frac{\bm{S}_t}{n_t} \big)^{-1} - \frac{1}{p} (\bm{\mu} - \bm{\mu}_t)^\top\bm{\Sigma}^{-1}_t\big\} \bm{x}\mid  
\label{4.2.11}
\end{align}
The first term on the right hand side converges in probability to zero by WLLN. The second term converges in probability to zero by WLLN, Slutsky's theorem and continuous mapping theorem. By Cauchy-Schwarz inequality, the third term is bounded above by $|| \big\{\frac{n}{n_t}(\bar{\bm{X}} - \bar{\bm{X}}_t)^\top \big(\frac{\bm{S}_t}{n_t} \big)^{-1} - \frac{1}{p} (\bm{\mu} - \bm{\mu}_t)^\top\bm{\Sigma}^{-1}_t\big\}||_2 \Big\{\sup\limits_{\bm{x} \in \text{supp}(\bm{X}_i)}||\bm{x}||_2 \Big\}$. Since $||\bm{x}||_2$ is bounded, this term converges in probability to zero. This proves part (a) of the Theorem. Part (b) can be proved similarly by switching the role of treatment and control group.


\subsection{Proof of Theorem 2} 
Consider the first term of the MRI estimator $\sum_{i:Z_i=1}w^{\scriptscriptstyle \text{MRI}}_i Y^{\text{obs}}_i$. By standard OLS theory, when $m_1(\bm{x})$ is linear on $\bm{x}$ we have
\begin{equation}
    \sum_{i:Z_i=1}w^{\scriptscriptstyle \text{MRI}}_i Y^{\text{obs}}_i = \hat{\beta}_{0t} + \hat{\bm{\beta}}^\top_{1t} \bar{\bm{X}} \xrightarrow[n \to \infty]{P} {E}(m_1(\bm{X}_i)) = {E}(Y_i(1)).
    \label{4.3.1}
\end{equation}
Similarly, when $m_0(\bm{x})$ is linear on $\bm{x}$,
\begin{equation}
    \sum_{i:Z_i=0}w^{\scriptscriptstyle \text{MRI}}_i Y^{\text{obs}}_i = \hat{\beta}_{0c} + \hat{\bm{\beta}}^\top_{1c} \bar{\bm{X}} \xrightarrow[n \to \infty]{P} {E}(m_0(\bm{X}_i)) = {E}(Y_i(0)).
    \label{4.3.2}
\end{equation}
Equations \ref{4.3.1} and \ref{4.3.2} prove part (iii) of the Theorem.

Now, let $p,\bm{\mu}, \bm{\mu}_t, \bm{\Sigma}_t$ be as in the proof of Theorem 1.
\begin{align}
\sum_{i:Z_i=1}w^{\scriptscriptstyle \text{MRI}}_i Y^{\text{obs}}_i &= \bar{Y}_t + (\bar{\bm{X}} - \bar{\bm{X}}_t)^\top \big( \frac{\bm{S}_t}{n_t}\big)^{-1} \Big\{\frac{1}{n_t}\sum_{i:Z_i=1}(\bm{X}_i- \bar{\bm{X}}_t)Y^{\text{obs}}_i \Big\} \nonumber\\
& \xrightarrow[n \to \infty]{P} {E}(Y^{\text{obs}}_i\mid Z_i=1) + (\bm{\mu} - \bm{\mu}_t)^\top \bm{\Sigma}^{-1}_t \Cov(\bm{X}_i, Y^{\text{obs}}_i\mid Z_i=1) ,
\label{4.3.3}
\end{align}
where the above convergence holds by a combination of WLLN, Slutsky's theorem and continuous mapping theorem. Under unconfoundedness, ${E}(Y^{\text{obs}}_i\mid Z_i=1) = \frac{1}{p}{E}(m_1(\bm{X}_i)e(\bm{X}_i))$,  Similarly, $\Cov(\bm{X}_i, Y^{\text{obs}}_i\mid Z_i=1)= \frac{1}{p} \big( {E}(\bm{X}_i m_1(\bm{X}_i)e(\bm{X}_i)) - \bm{\mu}_t {E}(m_1(\bm{X}_i)e(\bm{X}_i)) \big)$. This implies,
\begin{align}
{E}(Y^{\text{obs}}_i\mid Z_i=1) + (\bm{\mu} - \bm{\mu}_t)^\top \bm{\Sigma}^{-1}_t \Cov(\bm{X}_i, Y^{\text{obs}}_i\mid Z_i=1) \nonumber\\
= {E}\Big(\frac{m_1(\bm{X}_i)e(\bm{X}_i)}{p} \big\{1+ (\bm{\mu} - \bm{\mu}_t)^\top \bm{\Sigma}^{-1}_t(\bm{X}_i - \bm{\mu}_t) \big\}    \Big)
\label{4.3.4}
\end{align} 
Now, if $e(\bm{x})$ is inverse-linear on $\bm{x}$, by Equation \ref{4.2.1} in the proof of Theorem 1, we have $\frac{1}{e(\bm{x})} = \frac{1}{p} \big\{1 + (\bm{x} - \bm{\mu}_t)^\top \bm{\Sigma}^{-1}_t (\bm{\mu} - \bm{\mu}_t)   \big\}$. From Equation \ref{4.3.4}, we get ${E}(Y^{\text{obs}}_i\mid Z_i=1) + (\bm{\mu} - \bm{\mu}_t)^\top \bm{\Sigma}^{-1}_t \Cov(\bm{X}_i, Y^{\text{obs}}_i\mid Z_i=1) = {E}(m_1(\bm{X}_i)) = {E}(Y_i(1))$. Therefore, if $e(\bm{x})$ is inverse-linear, we have
\begin{equation}
\sum_{i:Z_i=1}w^{\scriptscriptstyle \text{MRI}}_i Y^{\text{obs}}_i \xrightarrow[n \to \infty]{P} {E}(Y_i(1)).    
\label{4.3.5}
\end{equation}
Similarly, if $1-e(\bm{x})$ is inverse-linear, we have
\begin{equation}
\sum_{i:Z_i=0}w^{\scriptscriptstyle \text{MRI}}_i Y^{\text{obs}}_i \xrightarrow[n \to \infty]{P} {E}(Y_i(0)).    
\label{4.3.6}
\end{equation}
Equations \ref{4.3.2} and \ref{4.3.5} prove consistency of the MRI estimator under condition (i) of the Theorem. Equations \ref{4.3.1} and \ref{4.3.6} prove consistency under condition (ii). When $e(\bm{x})$ is constant, then both $e(\bm{x})$ and $1-e(\bm{x})$ can be regarded as inverse-linear on $\bm{x}$ and hence consistency under condition (iv) follows from Equations \ref{4.3.5} and \ref{4.3.6}. Finally, when $p^2 \Var(\bm{X}_i\mid Z_i=1) = (1-p)^2 \Var(\bm{X}_i\mid Z_i=0)$, the URI and MRI estimator are asymptotically equivalent. Hence, consistency under (v) holds by similar argument as in the URI case (see the proof of Theorem \ref{dr_uri}.

\subsection{Proof of Proposition 4}
For $g \in \{t,c\}$, let $\underline{\tilde{\bm{X}}}_g$ be the design matrix in treatment group $g$. Also, let $\underline{\tilde{\bm{X}}}$ be the design matrix in the full-sample. 
We consider the MRI approach first. Without loss of generality, we compute the sample influence curve for a treated unit $i$. Since the two regression models in MRI are fitted separately, the SIC for unit $i$ for the MRI estimator of the ATE is the same as that for the MRI estimator of $\hat{{E}}(Y(1))$.
Let $\hat{\bm{b}}_t: = (\hat{\beta}_{0t},\hat{\bm{\beta}}^\top_{1t})^\top$ be the estimated vector of coefficients in the regression model in the treatment group. Also, let $\hat{\bm{b}}_{(i)t}$ be the corresponding estimated vector of coefficients when the model is fitted excluding unit $i$. It follows that (see \citealt{cook1982residuals}, Chapter 3),
\begin{equation}
    \hat{\bm{b}}_t - \hat{\bm{b}}_{(i)t} = (\underline{\tilde{\bm{X}}}^\top_t \underline{\tilde{\bm{X}}}_t)^{-1} \tilde{\bm{X}_i}\frac{e_i}{1-h_{ii,t}},
    \label{t5.1.1}
\end{equation}
where $\tilde{\bm{X}_i} = (1,\bm{X}^\top_i)^\top$. Denote $\tilde{\bar{\bm{X}}} = (1,\bar{\bm{X}}^\top)^\top$. Since $\hat{{E}}(Y(1)) = \tilde{\bar{\bm{X}}}^\top \hat{\bm{b}}_t$. Therefore, the SIC of unit $i$ is given by
\begin{equation}
    \text{SIC}_i = (n_t-1)(\tilde{\bar{\bm{X}}}^\top \hat{\bm{b}}_t - \tilde{\bar{\bm{X}}}^\top\hat{\bm{b}}_{(i)t}) = (n_t-1)\tilde{\bar{\bm{X}}}^\top(\underline{\tilde{\bm{X}}}^\top_t \underline{\tilde{\bm{X}}}_t)^{-1} \tilde{\bm{X}_i}\frac{e_i}{1-h_{ii,t}}
    \label{t5.1.2}
\end{equation}
We observe that $\sum_{i:Z_i=1} w^{{\scriptscriptstyle \text{MRI}}}_i Y^{\text{obs}}_i = \tilde{\bar{\bm{X}}}^\top \hat{\bm{b}}_t = \tilde{\bar{\bm{X}}}^\top (\underline{\tilde{\bm{X}}}^\top_t \underline{\tilde{\bm{X}}}_t)^{-1} \underline{\tilde{\bm{X}}}^\top_t \bm{y}_t$, where $\bm{y}_t$ is the vector of outcomes in the treatment group. So we can alternatively express the MRI weights in the treatment group as $w^{{\scriptscriptstyle \text{MRI}}}_i = \tilde{\bar{\bm{X}}}^\top(\underline{\tilde{\bm{X}}}^\top_t \underline{\tilde{\bm{X}}}_t)^{-1} \tilde{\bm{X}_i}$. It follows from Equation \ref{t5.1.2} that,
\begin{equation}
    \text{SIC}_i = (n_t-1)\frac{e_i}{1-h_{ii,t}}w^{{\scriptscriptstyle \text{MRI}}}_i.
    \label{t5.1.3}
\end{equation}
This completes the proof for MRI.

Let $\bm{l} = (0,0,...0,1) \in \mathbb{R}^{k+2}$. Consider the URI regression model $Y^{\text{obs}}_i = \beta_0 + \bm{\beta}^\top_1 \bm{X}_i + \tau Z_i + \epsilon_i$. Similar to the MRI case, let $\hat{\bm{b}}$ (respectively, $\hat{\bm{b}}_{(i)}$) be the vector of regression coefficients when the regression model is fitted using all the units (respectively, all excluding the $i$th unit). By similar calculations as before, it follows that, 
\begin{equation}
    \hat{\bm{b}} - \hat{\bm{b}}_{(i)} = (\underline{\tilde{\bm{X}}}^\top \underline{\tilde{\bm{X}}})^{-1} \tilde{\bm{X}_i}\frac{e_i}{1-h_{ii}},
    \label{t5.1.4}
\end{equation}
Now the URI estmimator $\hat{\tau}^{\text{OLS}}$ can be expressed as,
\begin{equation}
    \hat{\tau}^{\text{OLS}} = \bm{l}^\top \hat{\bm{b}} = \bm{l}^\top (\underline{\tilde{\bm{X}}}^\top \underline{\tilde{\bm{X}}})^{-1} \tilde{\bm{X}}^\top\bm{y}. 
    \label{t5.1.5}
\end{equation}
Since $\hat{\tau}^{\text{OLS}} = \sum_{i:Z_i=1} w^{{\scriptscriptstyle \text{URI}}}_i Y^{\text{obs}}_i - \sum_{i:Z_i=0} w^{{\scriptscriptstyle \text{URI}}}_i Y^{\text{obs}}_i$, we can alternatively express the URI weight of unit $i$ as $w^{{\scriptscriptstyle \text{URI}}}_i = (2Z_i-1)\tilde{\bm{X}}_i (\underline{\tilde{\bm{X}}}^\top \underline{\tilde{\bm{X}}})^{-1} \bm{l}$. Therefore, the sample influence curve of unit $i$ is given by
\begin{align}
   \text{SIC}_i = (n-1)(\bm{l}^\top \hat{\bm{b}} - \bm{l}^\top\hat{\bm{b}}_{(i)}) = (n-1)\bm{l}^\top(\underline{\tilde{\bm{X}}}^\top \underline{\tilde{\bm{X}}})^{-1} \tilde{\bm{X}_i}\frac{e_i}{1-h_{ii}} \nonumber\\
   = (n-1) \frac{e_i}{(1-h_{ii})}(2Z_i-1)w^{\scriptscriptstyle \text{URI}}_i.
    \label{t5.1.6} 
\end{align}
This completes the proof for URI.




\end{document}